\documentclass{aa}
\usepackage{natbib,twoopt}
\usepackage{amsmath}

\usepackage{color}
\usepackage[breaklinks=true]{hyperref} %% to avoid \citeads natbib format fothe density)) ApJ
\makeatletter
  \newcommandtwoopt{\intheteads}[3][][]{\href{http://adsabs.harvard.edu/abs/#3}%
    {\def\hyper@linkstart##1##2{}%
     \let\hyper@linkend\@empty\citealp[#1][#2]{#3}}}
  \newcommandtwoopt{\citepads}[3][][]{\href{http://adsabs.harvard.edu/abs/#3}%
    {\def\hyper@linkstart##1##2{}%
     \let\hyper@linkend\@empty\citep[#1][#2]{#3}}}
  \newcommandtwoopt{\citetads}[3][][]{\href{http://adsabs.harvard.edu/abs/#3}%
    {\def\hyper@linkstart##1##2{}%
     \let\hyper@linkend\@empty\citet[#1][#2]{#3}}}
  \newcommandtwoopt{\citeyearads}[3][][]%
    {\href{http://adsabs.harvard.edu/abs/#3}
    {\def\hyper@linkstart##1##2{}%
     \let\hyper@linkend\@empty\citeyear[#1][#2]{#3}}}
\makeatother

\usepackage{graphicx}
\usepackage{txfonts}
\begin{document}

\title{Radial properties of dust in galaxies: Comparison between observations and isolated galaxy simulations}

\authorrunning{S.A. van der Giessen et al.}

   \author{S.A. van der Giessen
          \inst{1,2}
          \and K. Matsumoto \inst{1,3,4}
          \and M. Relano \inst{2,5}
                  \and I. De Looze \inst{1}
                  \and L. Romano \inst{6,7,8}
            \and H. Hirashita \inst{9,10}
            \and K. Nagamine \inst{10,11,12,13}
            \and M. Baes\inst{1}
            \and M. Palla \inst{1,14,15}
            \and K.C. Hou \inst{16}
            \and C. Faesi \inst{17}
         }
\institute{Sterrenkundig Observatorium, Ghent University, Krijgslaan 281 - S9, 9000 Gent, Belgium
        \and 
            Dept. Fisica Teorica y del Cosmos, Universidad de Granada, Spain
        \and
            Department of Physics, Graduate School of Science, The University of Tokyo, 7-3-1 Hongo, Bunkyo-ku, Tokyo 113-0033, Japan
        \and
            Institute of Space and Astronautical Science, Japan Aerospace Exploration Agency, 3-1-1 Yoshinodai, Chuo-ku, Sagamihara, Kanagawa 252-5210, Japan
        \and
            Instituto Universitario Carlos I de Física Te\'{o}rica y Computacional, Universidad de Granada, 18071, Granada, Spain
        \and
            Universit\"{a}ts-Sternwarte, Fakult\"{a}t f\"{u}r Physik, Ludwig-Maximilians-Universit\"{a}t M\"{u}nchen, Scheinerstr. 1, D-81679 M\"{u}nchen, Germany
        \and
            Max-Planck-Institut f\"{u}r extraterrestrische Physik, Giessenbachstr. 1, D-85741 Garching, Germany
        \and
            Excellence Cluster ORIGINS, Boltzmannstr. 2, D-85748 Garching, Germany
        \and
            Institute of Astronomy and Astrophysics, Academia Sinica, Astronomy-Mathematics Building, No. 1, Section 4, Roosevelt Road, Taipei 106216, Taiwan
        \and
             Theoretical Astrophysics, Department of Earth and Space Science, Osaka University, 1-1 Machikaneyama, Toyonaka, Osaka 560-0043, Japan
        \and
             Theoretical Joint Research, Forefront Research Center, Osaka University, 1-1 Machikaneyama, Toyonaka, Osaka 560-0043, Japan
        \and
            Kavli IPMU (WPI), UTIAS, The University of Tokyo, Kashiwa, Chiba 277-8583, Japan
        \and
            Department of Physics and Astronomy, University of Nevada, Las Vegas, 4505 S. Maryland Pkwy, Las Vegas, NV 89154-4002, USA 
        \and
            Dipartimento di Fisica e Astronomia “Augusto Righi”, Alma Mater Studiorum, Università di Bologna, Via Gobetti 93/2, 40129 Bologna, Italy
        \and 
            INAF – Osservatorio di Astrofisica e Scienza dello Spazio di Bologna, Via Gobetti 93/3, 40129 Bologna, Italy
        \and
            Physics Department, Ben-Gurion University of the Negev, Be’er-Sheva 84105, Israel
        \and
            Department of Physics, University of Connecticut, Storrs, CT,
            06269, USA\\
\email{stefan.stefananthonyvandergiessen@ugent.be}
             }
   \date{Accepted: October 12th 2024}

\abstract{We study the importance of several processes that influence the evolution of dust and its grain size distribution on spatially resolved scales in nearby galaxies. Here, we compiled several multi-wavelength observations for the nearby galaxies NGC\,628 (M74), NGC\,5457 (M101), NGC\,598 (M33), and NGC 300. We  applied spatially resolved spectral energy distribution (SED) fitting to the latest iteration of infrared data to get constraints on the galaxy dust masses and the small-to-large grain abundance ratio (SLR). We separated each galaxy into radial rings and obtained the radial profiles of the properties mentioned above. For comparison, we took the radial profiles of the stellar mass and gas mass surface density for NGC\,628 combined with its metallicity gradient in the literature to calibrate a single-galaxy simulation using the \texttt{GADGET4-OSAKA} code. The simulations include a parametrization to separate the dense and diffuse phases of the ISM where different dust-evolution mechanisms are in action. We find that our simulation can reproduce the radial profile of dust mass surface density but overestimates the SLR in NGC\,628. Changing the dust-accretion timescale has little impact on the dust mass or SLR, as most of the available metals are accreted onto dust grains at early times ($<3$\,Gyr), except in the outer regions of the galaxy where the metallicity is below $2\times10^{-3}$. This suggests we can only constrain the accretion timescale of galaxies at extremely low metallicities where accretion still competes with other mechanisms controlling the dust budget. The overestimation of the SLR likely results from (i) overly efficient shattering processes in the diffuse interstellar medium (ISM), which were calibrated to reproduce Milky Way-type galaxies and/or (ii) our use of a diffuse and dense gas density subgrid model that does not entirely capture the intricacies of the small-scale structure present in NGC\,628. We conclude that future modeling efforts will need to focus on improving the subgrid recipes to mimic the multi-phase gas distribution in galaxies before the efficiency of dust evolution processes can be calibrated for galaxies other than the Milky Way.}

\maketitle

%------------------------------------------------------------
\section{Introduction}\label{Sec:Intro}

The existence of interstellar dust has been accepted ever since the first observation of obscured stellar emission in interstellar clouds \citep[e.g.,][]{Schal29, Trum30}. Even with dust only making up 1$\mathrm{\%}$ of the  mass of the interstellar medium (ISM), it absorbs, scatters, and reradiates a large portion of interstellar light depending on the dust-grain size distribution and dust material \citep[e.g.,][]{Driver07, Jones17, Bian18, Gall22}. Dust coexists with molecular gas such as $\mathrm{H_{2}}$ and CO, as the radiative interaction allows dust to shield molecules from interstellar ultraviolet light, which would otherwise dissociate the molecules \citep[e.g.,][]{Holl71, Dwek87, Hira17}. These molecules help with star formation as they allow the gas to cool and create high-density regions \citep[e.g.,][]{Kenn07, Bigi11, Ler13, Schruba11}. These radiative and chemical roles of dust in astrophysical processes make the study of the origin and evolution of dust a crucial undertaking.

Part of the dust originates as a byproduct of the evolution of stars. Asymptotic giant branch (AGB) stars contribute to the dust budget through processes such as dredge-up, in which part of the material from deeper layers gets released by convective streams to the surface and from there gets ejected by stellar winds. Heavier stars explode in supernovae (SNe) and instead eject  heavy metals from which dust grains can condense after the expanding ejecta have sufficiently cooled down \citep{Mats09, Temim12, Ven14, Looze17, Looze19, Shahba23}. SN shocks reprocess or even destroy dust grains ---depending on the type and size--- via processing through the blastwave \citep{Sla15, Hu19, Kirch22} or reverse shock \citep{Mice16, Boc16, Sla20, Kirch19, Kirch23, Kirch24a, Kirch24b}.
Once the stars release the dust, it interacts with the ISM. The metals released by stars or existing in the ISM can accrete onto the dust grains, causing grains to grow in mass. Dust grains also interact with each other via collisions, and could consequently coagulate into larger grains or shatter into smaller pieces \citep[e.g.,][]{Jones96, Hira09, Orm09, Aoy17}. Dust also interacts with thermally heated gas particles through energetic processes such as SN shock waves, and eventually gets destroyed. This interaction, called sputtering, causes a return of grain material back to the gas phase \citep[e.g.,][]{Barl78, Hira15, Hu19}.

The grain size distribution provides clues as to the balance of various dust formation and destruction processes. A viable way of deriving the grain size distribution from observations is to use spectral energy distributions (SEDs). The shape of the dust SED depends on the grain size, the relative contribution of the different dust species, and the dust temperature (assuming that the radiation field heats the dust) \citep{Draine07, Jones17}.
Thus, the methodology of extracting the grain size distribution from the SED heavily relies on the assumed dust model, and the radiation field heating the dust particles. This approach also often requires other observed galaxy properties acting as tracers for the different evolution mechanisms \citep{Gall21, Gall22}. 
For instance, each dust model proposes a different grain size distribution with varying separations between grain species to describe the dust emission features in the IR \citep[e.g.,][]{Des90, Jones17, Hens23}.

Separating dust into two size regimes has proven to be a good approximation, allowing authors to reproduce the main properties of the dust emission and extinction while saving computation time \citep[][]{Hira15}. This two-size approximation was successfully implemented in an isolated single galaxy simulation with the \texttt{GADGET3-OSAKA} code \citep{Aoy17} and was tested by evaluating the extinction curves as a function of time and galactic radius \citep{Hou17, Aoy20}.
Furthermore, \citet{Aoy18}, \citet{Hou19}, and \citet{Aoy19}  implemented and applied the two-size approximation to cosmological simulations and examined the dust evolution in a cosmological context. 
\cite{Rela20, Rela22} compared the dust SEDs inferred from observations with the above simulations. These authors successfully linked the SED variations with the individual mechanisms that affect the dust evolution. For a sample of nearby galaxies, 
\citet{Remy14} investigated  how the total dust mass varies with the metallicity $Z$, stellar mass ${M_{*}}$, star formation rate SFR, and specific star formation rate sSFR using dust emission in the IR by estimating the dust mass through models that include a combination of polycyclic aromatic hydrocarbons (PAHs), graphite, and silicates to describe the dust emission in the IR. These authors showed that the ratio of the total dust to gas mass, or the dust-to-gas ratio (DGR), increases strongly with $Z$, whereas it weakly correlates with ${M_{*}}$, SFR, and sSFR. The correlation is not equally strong across all $Z$, as the dwarf galaxies at lower $Z$ show greater variation in the DGR. 
When compared to chemical evolution models presented in \citet{Asan13} and \cite{Zhub14}, it became evident that dust growth by accretion is needed to match the observations. \citet{DeVis17} investigated this further and studied variations in the ratio between the total dust mass and the total amount of metals, the dust-to-metal mass ratio (DZR)\footnote{We consider here the total amount of metal in the gas phase of the ISM. The ratio changes insignificantly when we add the dust to the total amount of metal in the ISM.} , in a different sample of nearby galaxies in the DustPedia catalog \citep{Davies17}. \cite{DeVis17} concluded that both DGR and DZR increase with metallicity, hinting again at a variation of dust growth efficiency with metallicity, which was reproduced again in \citet{Gall21}. These scaling relations for the dust still hold on local scales for galaxies, as shown by four independent studies, namely \citet{Chiang18}, \citet{Casa22}, \citet{RomanDuval22}, and \citet{Clark23}, which all demonstrate that the DGR still increases with metallicity on scales of 3-4 kpc for different samples of nearby galaxies using different dust models.

In addition to studying variation in the total dust mass with physical galaxy properties, it is possible to study the scaling relations of grains with radii smaller than $0.015\,\mu m$ relative to larger grains, the small-to-large grain ratio (SLR), as this ratio contains information on the processes that affect the dust size distribution. \citet{Rela20} showed that the dust masses, DGR, and SLR derived from observations on spatially resolved scales in a set of three nearby galaxies, NGC\,628 (M74), NGC\,5457 (M101), and NGC\,598 (M33), agree with an isolated galaxy simulation of a Milky Way-like galaxy presented in \citet{Aoy17}. Based on this agreement, \citet{Rela20} concluded that coagulation becomes efficient in the centers of these galaxies probably because of the high molecular gas mass and metallicity. Still further from the center, these authors note an increase in the SLR as coagulation becomes less efficient and balances with small-grain-formation processes such as accretion and shattering.
In the outermost part of the galaxies, interstellar processing of dust becomes inefficient, and so the most efficient dust evolution channel is stellar production, which explains the observed decrease in the SLR. A follow-up of this study on a large sample of galaxies in the nearby Universe shows a significant fraction of galaxies with high integrated SLRs \citep{Rela22}. These high integrated values derived from observations are reproduced in simulations of an individual Milky Way-type galaxy that implement the diffusion of metals \citep{Roma22b}. These latter authors find that diffusion will transport large grains formed by stellar sources and coagulation in the dense ISM into the diffuse ISM, where the dust grains shatter more efficiently and produce high SLRs. 

For any galaxy simulation with a dust model, an important step is to replace the two-size approximation with a detailed grain size distribution for a more detailed view.
For this purpose, we use the \texttt{GADGET4-OSAKA} simulation \citep{Roma22a,Roma22b,Matsu24} ---which uses a dust evolution scheme for grains with 30 different size bins covering the grain size distribution between $\mathrm{3\,\AA}$ and $\mathrm{10\,\mu m}$ rather than only two-grain sizes--- to constrain the evolution of a single galaxy. We calibrate the stellar and gas properties of the simulations to reproduce one of the most well-studied nearby galaxies from \citet{Rela20}, NGC\,628. We investigate the efficiencies of dust evolution mechanisms such as dust grain growth and destruction, which depend highly on the grain sizes, by running simulations with different dust accretion timescales and parameter values controlling the dense gas fraction. We compare observable properties, such as the DZR and SLR, with these different simulations to understand the conditions that make the dust properties agree with NGC\,628 and how they should change to agree with other nearby galaxies. In particular, we compare the radial distributions of the simulations with observations and explain how the  radial variations of the physical conditions affect the dust-evolution mechanisms across the galaxy disk. We use observations of three other galaxies from similar studies, NGC\,5457, NGC\,598, and NGC\,300, to investigate potential variations in present-day dust properties and how the dust has been impacted by different evolutionary processes.

In a companion paper, \citet{Matsu24} performed \texttt{GADGET4-OSAKA} simulations to produce galaxies that resemble the Milky Way and NGC\,628 and post-processed the simulation snapshots with the radiative transfer code SKIRT \citep{Baes11, Camps15, Camps20}. They study how the PAH mass fraction varies with metallicity and how it is affected by other dust-processing mechanisms. In the present paper, we focus on the relative contribution of small and large grains, where small grains are considered to be PAHs and very small carbonaceous dust grains. We analyze how the SLR changes within the simulated galaxy disk and compare the results from simulations with observations. The details of the \texttt{GADGET4-OSAKA} simulations used in this paper are described in \citet{Matsu24}, as are the analyses of the PAH mass fraction. 

The structure of this paper is as follows.
We describe both the observations and the simulation data in Sect. \ref{Sec:Data}, and present the methodology for comparison in Sect. \ref{Sec:Method}.
We present the comparisons of simulated and observed dust scaling relations in Sect. \ref{Sec:Results},
which is followed by a discussion in Sect. \ref{Sec:Disq}. We finally conclude in Sect. \ref{Sec:Conq}. In this study, we assume a solar metallicity $Z_{\odot}$ of 0.0134 \citep{Asp09}.

%------------------------------------------------------------
\section{Data}\label{Sec:Data}
\subsection{Observational data}\label{Sec:Obs}
We selected nearby star-forming spiral galaxies from \citet{Rela20} and similar studies. Our galaxy sample includes the nearby star-forming spiral galaxies from \citet{Rela20}, namely NGC\,628, NGC\,5457, NGC\,598, and we add NGC\,300, a galaxy with similar physical properties to NGC\,598. Infrared data are available for these galaxies, which allow a spatially resolved study at spatial scales of \~2.7 kpc, therefore allowing resolution of the spiral structure within their discs. These galaxies have different physical properties, such as size and morphological type (see Tables~\ref{tab:Rings} and~\ref{tab:Prop}). In particular, they have different molecular gas mass contents and central metallicities, which allows us to explore how these two properties affect the dust evolution mechanisms. The four galaxies make for an interesting case study to see whether variations in the dust properties only depend on dust-related parameters, which we can control for in the simulations.

\begin{table*}[t]
\caption{Central coordinates, inclination, position angle, distance, and $R_{25}$ for the four galaxies in this study.}
\label{tab:Rings}
\resizebox{\linewidth}{!}{%
\begin{tabular}{l|l|l|l|l|l|l|l|l}
NGC reference & Messier reference   & Morphology type   & RA ($^{\circ}$)   & DE ($^{\circ}$)  & Inclination ($^{\circ}$)  & PA ($^{\circ}$)   & Distance (Mpc)    & $\mathrm{R_{25}}$(kpc)                \\\hline 
NGC\,628         & M74         & SA(s)c            & 24.17             & 15.78             & 5                         & 12                & 9.84              & 14.9      \\
NGC\,5457         & M101                  & SAB(rs)cd         & 210.80            & 54.35             & 18                        & 39                & 6.7               & 22.2        \\
NGC\,598         & M33                   & SA(s)cd           & 23.46             & 30.66             & 57                        & 23                & 0.84              & 6.85      \\
NGC\,300         & -                  & SA(s)d         & 13.72            & -37.68             & 39.8                        & 114.3                & 2.09               & 5.9        
\end{tabular}%
}
\end{table*}

\begin{table*}[t]
\caption{Total stellar mass, star formation rate, total gas mass, molecular gas mass, and 12+log(O/H) gradient for the four galaxies in this study.}
\label{tab:Prop}
\resizebox{\linewidth}{!}{%
\begin{tabular}{l|l|l|l|l|l}
NGC reference   &$\mathrm{log(M_{*}[M_{\odot}])}$ &$\mathrm{log(SFR[M_{\odot}\,yr^{-1}])}$ &$\mathrm{log(M_{HI}[M_{\odot}])}$ &$\mathrm{log(M_{H_{2}}[M_{\odot}])}$    & 12 + log(O/H)                     \\\hline 
NGC\,628         &10.34&0.24&9.70&9.08& $^{a}$8.835 - 0.485$\times$ r/$R_{25}$   \\
NGC\,5457              &10.3&0.53&10.15&9.5& $^{b}$8.716 - 0.832$\times$ r/$R_{25}$  \\
NGC\,598               &9.4&-0.40&9.15&8.52& $^{c}$8.50 - 0.382$\times$ r/$R_{25}$     \\
NGC\,300         &9.27&-0.82&9.32&7.67& $^{d}$8.57 - 0.41$\times$ r/$R_{25}$  

\end{tabular}%
}
\newline
\tablebib{a) \citet{Berg15}, b) \citet{Crox16}, c) \citet{Bres11}, d) \citet{TSC16}.}
\end{table*}

For this study, we use a collection of data from GALEX in the UV to IR from \textit{Spitzer}, WISE, and \textit{Herschel} to constrain the stellar and dust emission, as well as the SFR. For NGC\,628, NGC\,5457, and NGC\,300, the WISE, GALEX, and \textit{Herschel} data were obtained from the DustPedia collaboration \citep{Davies17, 2018A&A...609A..37C}. NGC\,598 is not part of DustPedia; therefore we use the updated scanamorphos-processed version\footnote{Scanamorphos post-processing \citep{2013PASP..125.1126R} enables us to efficiently subtract the low-frequency noise of \textit{Herschel} observations and create high-quality maps in the different observed bands.} of the \textit{Herschel} images from J. Chastenet et al. (in prep) and we downloaded the WISE images from the WISE Data Archive following the methodology presented in \citet{2018A&A...609A..37C} to build up the WISE mosaics. The \textit{Spitzer} images for all four galaxies were obtained from the LVL Survey \citep{2009ApJ...706..599L}.  We retrieved the GALEX UV images for NGC\,598 from the GALEX Data Archive \citep{2007ApJS..173..682M}.

The different observations do not have the same angular resolution, which is crucial when comparing at spatially resolved scales. We convolved and reprojected to the photometric band with the worst angular resolution to ensure this comparison. This corresponds to the angular resolution of $36"$ from the SPIRE $500\,\mathrm{\mu m}$, and a linear scale of 1720, 1170, 150, and 365\,pc for NGC\,628, NGC\,5457, NGC\,598, and NGC\,300, respectively. 
We removed the contamination from foreground stars in the UV and IR wavelength range separately. We followed the methodology described in \citet{2007ApJS..173..185G} to identify the foreground stars in the UV filters, which proved to be successful for 1000 nearby galaxies \citep{2007ApJS..173..185G}, as well as for M31 \citep{2014A&A...567A..71V}. For the IR images, we created a catalog of foreground stars based on the \textit{Spitzer} IRAC\,8\,$\mathrm{\mu m}$ images and the $3.6\,\mathrm{\mu m}-5.8\,\mathrm{\mu m}$ colors, following \citet{2009ApJ...703.1569M}. The foreground stars were replaced with random noise obtained from a local background surrounding the star. We furthermore confirmed that the WISE  22$\mathrm{\mu m}$ and \textit{Spitzer} 24$\mathrm{\mu m}$ band images were free of foreground stellar contamination by visually inspecting there was no emission at the location of the foreground stars.

The H\,{\sc{i}} 21cm and CO maps for the four galaxies come from different studies. We use the H\,{\sc{i}} 21cm map from The H\,{\sc{i}} Nearby Galaxies Survey \citep{Walt08} for NGC\,628 and NGC\,5457, \citet{Grat10} for NGC\,598, and \citet{Puch90} for NGC\,300. To describe the molecular gas mass content, we use the $\mathrm{^{12}CO(2-1)}$ maps from HERACLES \citep{Ler09, Ler13} for NGC\,628 and NGC\,5457, \citet{Drua14} for NGC\,598, and the ALMA CO(2-1) map from C. Faesi et al. (in prep) for NGC\,300. We convolved and re-gridded the H\,\textsc{i} and CO maps to make them align with the IR data set and converted the units into surface densities (M$_\odot$/pc$^2$) using Eq. 2 in \citet{DiTeo21} for H\,{\sc{i}} observations and Eq. 2 in \citet{Sand13} with a Milky Way $\alpha_{\rm CO}$ factor of 4.3 for the CO data \citep{Bola13}. The inclination of the galaxies shown in Table~\ref{tab:Rings} was used to convert observed fluxes into surface density units. 

For metallicities, we make use of the radial gradients in the literature constrained by the oxygen abundances in H\,\textsc{ii} regions and assume that we can neglect the variations along the rings with a fixed distance from the center, as studies showed these variations to be close to 1\% \citep[e.g.,][]{Kre19}. We use the metallicity gradient 
%for the radius normalized by $R_{25}$ (see Table~\ref{tab:Rings}) 
of \citet{Berg15} for NGC\,628, \citet{Crox16} for NGC\,5457, \citet{Bres11} for NGC\,598, and \citet{TSC16} for NGC\,300 (see Table \ref{tab:Prop}). 

Stellar mass maps were derived from the IRAC $\mathrm{3.4\,\mu m}$ and $\mathrm{4.5\,\mu m}$ images using the calibration provided in \citet{2012AJ....143..139E}. SFR maps were created using the combined FUV and total-IR (TIR) luminosities and the calibration given in \citet{2012ApJ...761...97M}. To obtain the TIR luminosity, we made use of the prescriptions provided in \citet{2011AJ....142..111B}, which are consistent with those derived in other studies \citep{Gala13}.

\subsection{Dust mass derivations}\label{Sec:dustmaps}
We used the classical \citet{1990A&A...237..215D} dust model, which consists of three different grain populations:  PAHs, very small grains (VSGs), and big silicate grains (BGs). The model assumes that PAHs, VSGs, and BGs correspond to grains with radii of ($0.4-1.2)\times 10^{-3}\,\mathrm{\mu m}$,
($1.2-15)\times 10^{-3}\,\mathrm{\mu m}$, and
$(15-110)\times 10^{-3}\,\mathrm{\mu m}$, respectively. The two-grain-size approximation proposed by \cite{Hira15} separates the grain population into small and large grains at the radius of $a=0.03$\,$\mathrm{\mu m}$. They justify this split based on the fact that when considering the dominant processes affecting small dust grains (accretion, shattering, and coagulation) within a full grain size distribution approach, the distribution starts to increase for radii smaller than 0.03\,$\mathrm{\mu m}$ \citep{Asan13b}. At $a=0.015$\,$\mathrm{\mu m}$, which is the separation between the small and large grains adopted here, the distribution has not arrived at the maximum yet. Therefore, we can assume that this value is also valid for separating small and large dust grains. A comparison of the grain size distribution adopted in the two-size approximation and a power-law distribution, which is the one assumed by \citet{1990A&A...237..215D}, is shown in Fig.A1 from \cite{Hira15}. The two-size grain distribution is able to reproduce the Milky Way dust extinction curve relatively well, as is the power law, with significant deviations only in the 2175\,$\AA$ bump. Therefore, we consider that the threshold of $a=0.015$\,$\mathrm{\mu m}$ adopted here is reasonable for separating small and large dust grains.

Using the \citet{1990A&A...237..215D} dust model enables a robust comparison with the previous work of \citet{Rela20}. However, unlike \citet{Rela20}, where only the pixels with high S/N fluxes in all the observed bands were fitted, here we improve on their observational analysis by extending the fit to cover a larger spatial area. We have designed a methodology to recover information from the outer parts of the galaxies where the low S/N makes the SED fitting less robust. Our new methodology shows a smooth radial decline of the dust mass surface density at larger galactocentric distances, while the  radial trend in the small-to-large grain ratio remains the same. We furthermore compare the dust mass and SLR obtained here with those derived by the more frequently used Themis dust model \citep{Jones17}, which was calibrated to the Milky Way diffuse ISM, similar to the model of \citet{1990A&A...237..215D}. This comparison, and a more detailed explanation of the Themis dust model, are presented in Appendix\,\ref{App:dustmethod}. The differences in dust masses increase with radial distance, with Themis dust masses being a factor 1.5 larger in the center of the galaxy, and similar at radii larger than 0.6\,$\rm R_{\rm 25}$. The radial trends for the SLR for both dust models are the same, but Themis dust models give lower ratios at all radial distances. We attribute these variations to differences in the dust grain model and we stress that these small deviations do not impact the main results of the present study.

\begin{figure}
 \includegraphics[width=\linewidth, clip=true, trim=0mm 1mm 15mm 10mm]{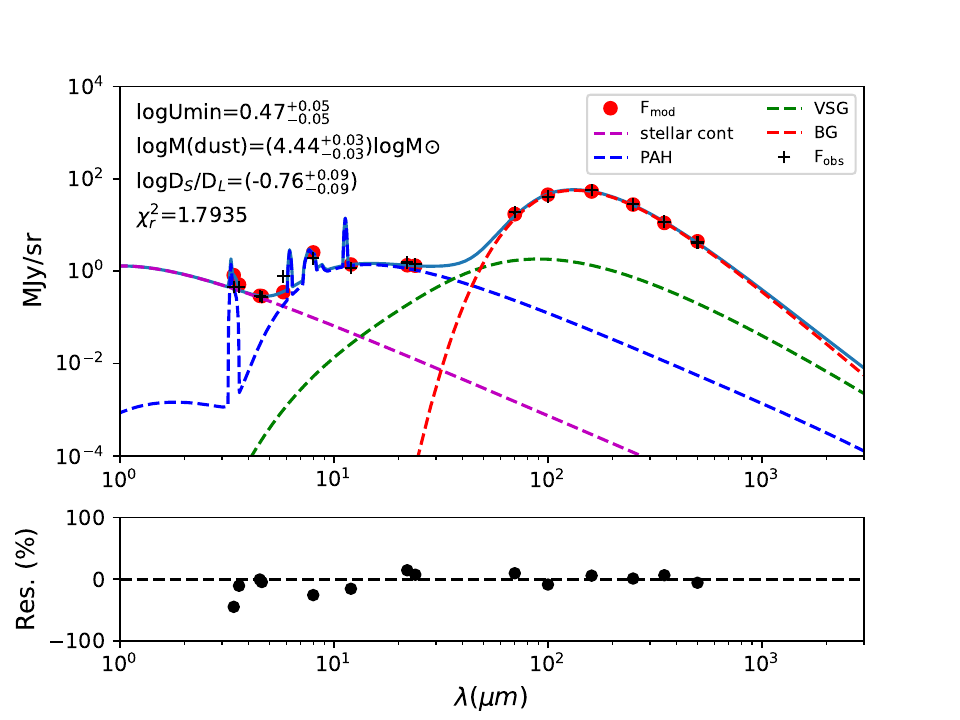}
 \caption{Example SED fitting in the disk of NGC\,628. The black crosses and red circles correspond to the observed and modeled fluxes, respectively. The purple dashed line represents a black body emission to account for the old stellar population, while the blue, green, and red dashed lines show the emission of the PAHs, VSGs, and BGs, respectively. The blue continuous line is the total emission. The panel below presents the residuals of the fit.} 
 \label{fig:sedfit}
\end{figure}

\begin{figure}[!ht]
    \centering
    \includegraphics[width=\linewidth, clip=true, trim=40mm 140mm 100mm 80mm]{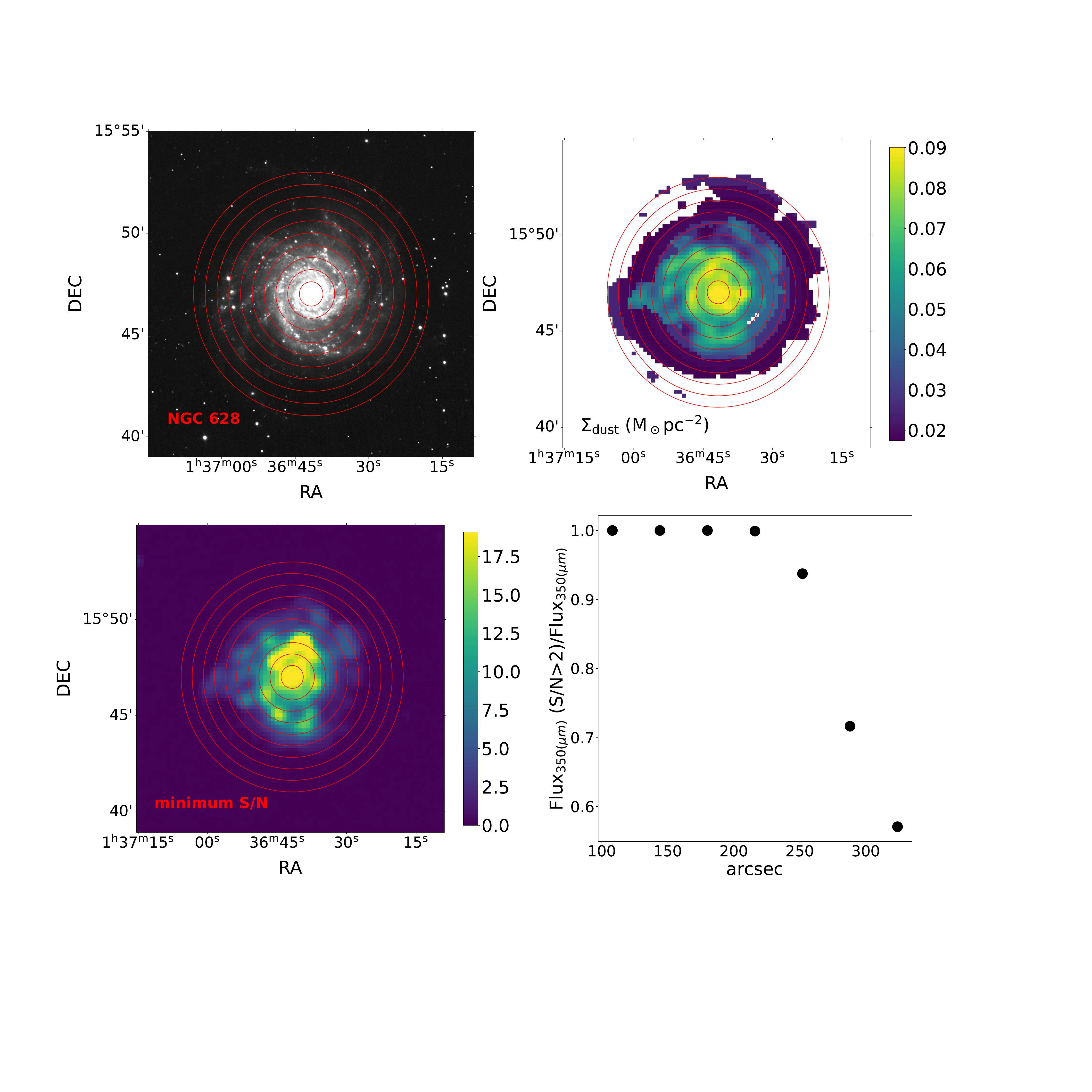}
    \caption{Spatially resolved images of NGC\,628. Top left: SDSS g-band image of NGC\,628 with concentric annuli of 36\,arcsec width. Top right: Dust-mass surface density of the whole disk.
    The color level indicates the dust surface density in units of M$_\odot$ pc$^{-2}$ as shown in the color bar.
    %We also show the same concentric annuli as shown in the top-left panel.
    Bottom left: S/N map corresponding to the minimum S/N of all the observed fluxes in each pixel. Bottom right: Total 350\,$\mathrm{\mu m}$ flux for the low-S/N pixels pixels with $\mathrm{S/N}>2$ in the SPIRE 350\,$\mu$m image with respect to the total flux from the low-S/N pixels in each ring.
    The same figure is shown in Appendix\,\ref{App:dustmethod} for the other galaxies in the sample.}
    \label{fig:dustmethod}
\end{figure}

We fit the observed IR SEDs at each location of the disk for our galaxy sample using nested sampling techniques \citep[{\tt dynesty}\footnote{\url{https://dynesty.readthedocs.io/en/sTable/}},][]{Spea20} to estimate the Bayesian posteriors and evidence in our Bayesian SED modeling method. The free parameters for our fitting strategy are the masses of the different components in the dust model: $M_{\rm PAH}$, $M_{\rm VSG}$, and $M_{\rm BG}$ in M$_{\odot}$, a scale factor for the %%ISRF
interstellar radiation field of the solar neighborhood \citep{1983A&A...128..212M}, $U_0$, and a parameter, $U_{\rm NIR}$, which takes into account the contribution of the old stellar population in the near-infrared (NIR) part of the SED, parameterized by a black body of temperature $T=5,000$\,K. We use flat priors in the logarithmic space for the distribution of the free parameters in the fit. To achieve a robust dust mass determination, we performed the fit to those pixels that have fluxes with a signal-to-noise  ratio (S/N) of greater than 2 {in all bands}, with the S/N in each pixel corresponding to the minimum S/N of all the observed bands. These pixels are called ``high-S/N pixels'' here, while those pixels that have at least one band with flux below $\mathrm{S/N}=2$ are called ``low-S/N pixels''. In the lower left panel of Fig. \ref{fig:dustmethod}, we present the minimum S/N map for NGC\,628, showing that most of the disk (up to the sixth ring depicted in the figure) has an S/N of greater than 2. At greater distances from the center, the spatial coverage in the map with high S/N values for each ring decreases, and the number of pixels with low S/N increases. In Fig.\,\ref{fig:sedfit} we show an example of a typical SED for a high-S/N pixel with the corresponding fit. The uncertainty of the best-fit 
parameters was obtained using the 16th and 84th percentiles of the resulting PDF.

As we compare our results with simulations for well-resolved galaxies, we need reliable dust mass estimates for the outer parts of the disk where the individual pixels do not fulfil our S/N criteria for all bands and most of the pixels correspond to what we call low-S/N pixels.
We apply a methodology to recover the dust emission in these outer regions, which heavily relies on the method presented in \citet{Chiang21} but has been modified according to our data. We created a mask based on the minimum S/N map presented in the bottom-left panel of Fig.\,\ref{fig:dustmethod} containing the pixels with observed fluxes above $\mathrm{S/N}=2$ in all bands. We then divided the galaxy into concentric elliptical annuli using the central position, position angle, and inclination of each galaxy (see Table\,\ref{tab:Rings}). For each ring, we work on those pixels that are outside our mask of the minimum S/N, the low S/N pixels. Assuming that the dust mass is well represented by the flux at 350\,$\mu$m, we consider for each ring only the low S/N pixels with a 350\,$\mu$m flux above a S/N of 2\footnote{We note that the minimum S/N mask is created by considering the pixels with fluxes that have a $\mathrm{S/N}>2$ in {{all} bands}. In this selection, we only take into account the emission in the 350\,$\mu$m band.} and integrate the emission for them to generate a characteristic SED of the low-S/N pixels. 
In the bottom-right panel of Fig. \ref{fig:dustmethod}, we show for each ring the total flux of the low-S/N pixels with $\mathrm{S/N}>2$ in the SPIRE 350\,$\mu$m image with respect to the total flux from the low-S/N pixels in each ring. In the inner rings, all the emission comes from the pixels showing $\mathrm{S/N}>2$ in the SPIRE 350\,$\mu$m band, but for the outer ones, the contribution of these pixels decreases as we go further out in the disk. We built the characteristic SED for the low-S/N pixels in each ring by integrating the emission in each band over the low-S/N pixels with $\mathrm{S/N}>2$ in the SPIRE 350\,$\mu$m image, and fit the observations with the same dust model and number of free parameters as we have used for the individual high-S/N pixels. The total dust mass derived from the fit in each ring is then distributed among the selected pixels in that particular ring. We combined the results with those from the individual high-S/N pixels and created a dust-mass map that extends to the outer parts of the galaxy where the emission in the 350\,$\mu$m band is observationally detected. The resulting dust-mass map is shown in the top-right panel of Fig.\,\ref{fig:dustmethod}.  

\subsection{Simulation data from \texttt{GADGET4-OSAKA}}
We use the results of a single isolated galaxy simulated with the smooth-particle hydrodynamics (SPH) code \texttt{GADGET4-OSAKA} \citep{Roma22a, Roma22b}, which is a modified version of the \texttt{GADGET-4} code \citep{Spring21}. Compared to its predecessor, \texttt{GADGET3-OSAKA} \citep{Aoy17,Aoy19,Shim19}, the code has been improved in many aspects, such as in the treatment of dust diffusion and evolution \citep{Roma22a, Roma22b}.  
For a detailed explanation of the isolated galaxy simulation by \texttt{GADGET4-OSAKA} used in this paper, we refer the reader to \citet{Matsu24}. In this section, we highlight the simulation algorithms that play a role in the dust evolution.

\subsubsection{Gas and stars}\label{Sec:Osaka_GAS}
In the \texttt{GADGET4-OSAKA} code, the hydrodynamical evolution of gas is solved with the SPH method with the stochastic conversion of gas particles into star particles once the densities of gas particles reach the density of $\mathrm{10 \ cm^{-3}}$. Here, we assume the star formation efficiency of $0.05$ to follow the Kennicutt-Schmidt law \citep{Kennicutt1998}.  Each star particle represents a star cluster with a Chabrier initial mass function \citep[IMF;][]{Chabr03}, and energy input (stellar feedback) and metal ejection occur subsequently.  The details of the stellar feedback model are also described in the above references. 
The code calculates the metal enrichment by SNe Ia, SNe II, and AGB stellar winds using the chemical evolution library CELib  \citep{Sait17}, tracking the elements H, He, C, O, Ne, Mg, Si, S, Ca, and Fe. Each source has its own delay time for metal injection into the ISM after the corresponding star-formation event. Ten percent of the metals generated by stars are assumed to be ejected as dust grains \citep{Inoue11, Kuo13}. 
The cooling of gas particles is calculated with the Grackle-3 library \citep{SmithBD17} along with the nonequilibrium chemistry, including the formation of $\mathrm{H_2}$ on dust grains \citep{Roma22b}. The photo-heating, photo-ionization, and photo-dissociation of UV background radiation based on \citet{Haardt12} is taken into account. Consequently, the gas particles can cool down to around the temperature of 10K.  

In terms of the dust evolution and $\mathrm{H_2}$ models, the simulations apply a subgrid model based on \citet{Aoy20} and \citet{Roma22a,Roma22b} to account for the different ISM phases in gas particles, as the simulations do not resolve gas with high density (${n_\mathrm{H}} = 10^3\,\rm cm^{-3}$), where dust grows and molecular hydrogen forms.
Thus, in this model, each gas particle is assumed to host dense clouds and diffuse gas. The density and temperature of the dense cloud is fixed at the density of ${n_\mathrm{H}} = 10^3\,\rm cm^{-3}$ and the temperature of $T=50$\,K, respectively. On the other hand, the density and temperature of the diffuse gas are estimated according to the pressure equilibrium between the diffuse gas and dense clouds in each gas particle.
The fraction of the dense clouds in each gas particle, $f_{\text{cloud}}(n_{\rm H})$, is estimated as follows:

\begin{equation}
    f_\mathrm{cloud} =\min (\alpha n_\mathrm{H},~1),
\end{equation}
where $n_\mathrm{H}$ is the density of gas particles, and $\alpha$ is a parameter controlling the mass fraction of the dense clouds in gas particles. The authors employed the $\alpha$ of $0.12$, so that the fraction of the dense clouds in their Milky-Way galaxy simulation was consistent with the estimated molecular gas fraction in nearby galaxies \citep{Cati18}. However, it is uncertain whether the fraction of the dense clouds follows the molecular gas fraction due to the variation of environments. We therefore vary $\alpha$ in the range from $0.12$ to $0.012$ to examine how this variation influences the molecular gas surface density ${\Sigma_\mathrm{H_{2}}}$ and the dust properties in our simulations.

\begin{figure}[!ht]
    \centering
    \includegraphics[width=0.95\linewidth]{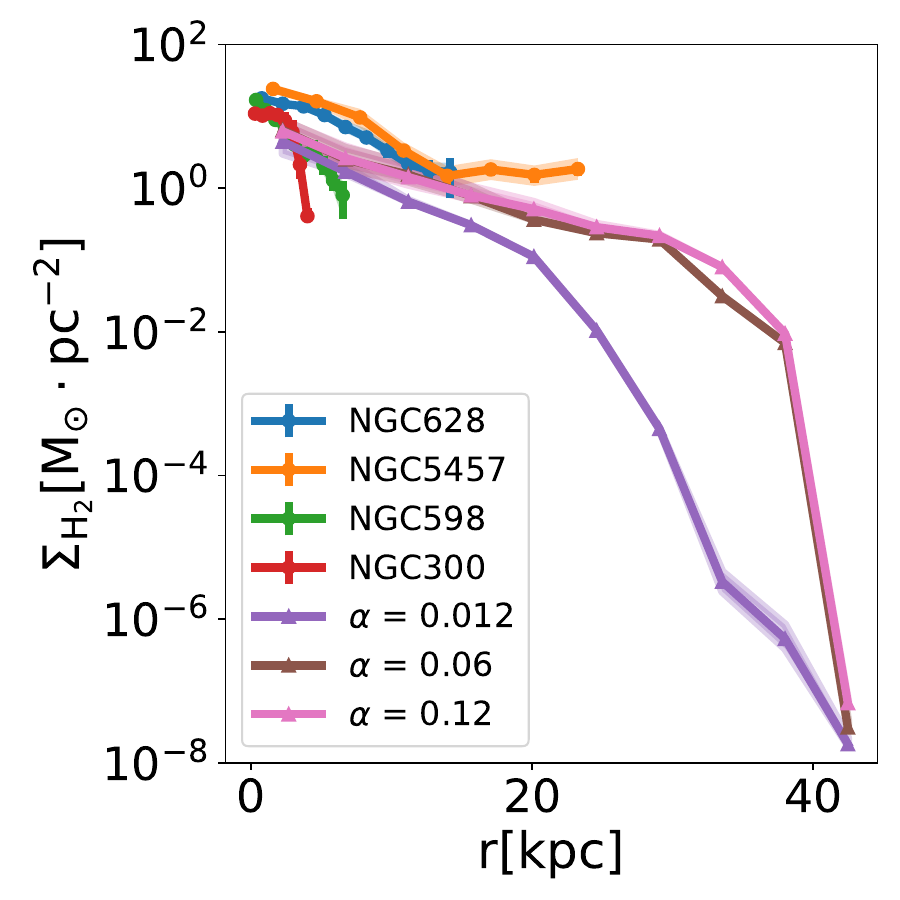}
    \caption{Radial profile of molecular-gas-mass surface density $\mathrm{\Sigma_{H_{2}}}$ in our simulation suite with three different values of $\alpha$ regulating the dense gas fraction. The lines indicate the median values in the radial bins with the uncertainties for observations. The shaded area is the $1\,\sigma$ deviation within each radial bin.}
    \label{fig:MH2_alpha}
\end{figure}

We compare the variation in ${\Sigma_\mathrm{H_{2}}}$ as observations use the ratio between the molecular gas and the total gas as a proxy for the dense gas fraction. Figure~\ref{fig:MH2_alpha} shows a comparison between the $\mathrm{\Sigma_{H_{2}}}$ radial profiles derived from our observations and those obtained by simulations with different $\alpha$ values.
The simulated $\mathrm{\Sigma_{H_{2}}}$ decreases consistently with radius for all $\alpha$ values considered here, as it does in the observed radial profiles. The molecular gas masses estimated from the simulations tend to be lower than those observed in NGC\,628, which is the galaxy we used to calibrate the simulations. The radial profiles for $\alpha$ values equal to 0.12 and 0.06 are quite similar; however for $\alpha=0.012$, the radial gradient of $\mathrm{\Sigma{_{H_{2}}}}$ is systematically stronger than for the two other $\alpha$ values. 
This implies that simulations with $\alpha = 0.012$ have a lower dense gas fraction at radii larger than 5 kpc. We keep this in mind when we analyze how the dust masses and SLR radial profiles vary with the different $\alpha$ values adopted.

\subsubsection{Dust model}\label{Sec:Data_Sim_Dust}
The simulations assume complete dynamical coupling between gas and dust; that is, no relative motion between them. 
We adopt the same dust evolution models as \citet{Roma22a}. These models, which are based on \citet{Asan13}, \citet{Aoy17}, \citet{Hira19}, and \citet{Aoy20}, take into account stellar dust production, destruction during the gas collapse for star formation or so-called astration, SN destruction, thermal sputtering, shattering, and coagulation in appropriate physical conditions of the ISM for each process. 
Stellar dust production follows a log-normal grain-size distribution with a median of 0.1 $\mathrm{\mu m}$ and a standard deviation of 0.47 dex \citep{Asan13b}.
SN destruction and thermal sputtering are modeled together as a general sputtering or dust destruction effect.
Shattering only happens in the diffuse and warm ISM where grain velocities are high enough for collisions to fragment the particles. Accretion and coagulation only take place in the dense and cold ISM where grain velocities are low, and the densities allow high encounter rates for dust--metal and dust--dust interactions.   The simulations assume a spherical dust grain using the silicate bulk density $s = 3.5\,\mathrm{g\,cm^{-3}}$ from \citet{Wein01} with possible sizes divided into 30 logarithmic bins spread between $3\times10^{-4}$ and 10 $\mathrm{\mu m}$. 

The accretion timescale $\tau_{\rm accr}$ is a parameter in dense clouds for the rate at which the grain mass grows through the sticking of gas-phase metals. 
The simulations calculate the accretion timescale given by 
\begin{equation}\label{eq:accretion}
    \tau_\mathrm{accr} = \frac{1}{3}\tau_{0}\left(\frac{a}{0.1\mathrm{\mu m}}\right)\left(\frac{Z}{Z_{\odot}}\right)^{-1}\left(\frac{n_\mathrm{H,\,cloud}}{10^{3}\mathrm{cm^{-3}}}\right)\left(\frac{T_\mathrm{cloud}}{10\mathrm{K}}\right)^{\frac{1}{2}},
\end{equation}
where $\tau_{0}$, $a$, and $Z$ are the normalized accretion timescale, grain radius $a$, and metallicity of gas particles, respectively. ${n_\mathrm{H, \,cloud}}$ and ${T_\mathrm{cloud}}$ are the density and temperature of dense clouds in gas particles.
Consequently, the dust density growth rate $\rho_\mathrm{d}/dt$ in each gas particle is calculated as follows \citep[see][]{Aoy20}:
\begin{equation}
    \left.\frac{d \rho_\mathrm{d}}{dt}\right|_\mathrm{accr} = \left(1 - \mathrm{DZR}\right)\frac{\rho_\mathrm{d}}{\tau_{accr}}.
\end{equation}
A higher DZR indicates that most gas-phase metals that could accrete onto dust grains are already locked up in grains, which reduces the amount of available metal for accretion. The simulations calculate this accretion timescale based on the total metal mass per simulation particle.

Although the simulations model dust evolution for a detailed grain-size distribution rather than employing the two-size approximation commonly used in other research \citep[e.g.,][]{Dubois2024}, several studies have shown that certain elements, such as iron, carbon, and silicon, are  more easily accreted onto dust grains than other elements \citep{Jones19, RomanDuval21, Konsta22, Konsta24}. This implies that the dust-depletion rate depends on the presence of specific chemical elements rather than on the metallicity alone. However, in the simulations, which only consider the dependence on the metallicity, the DZR reaches up to 0.9 at $t=10$ Gyr \citep{Matsu24}. We correct the simulated dust masses for this effect  during
post-processing by taking the ratio of elements that stick onto dust grains relative to the total elemental abundance in both gas and dust to reduce the obtained dust masses in the simulations. For this, we sum the element abundances relative to hydrogen mentioned in Table 1 of \citet{Jones19} for the Milky Way gas-phase ISM and dust, and then take the ratio between the two. This results in a metal ratio of 0.42. This ratio is then applied to correct the dust masses derived from the simulations from the element-dependent depletion. 
%------------------------------------------------------------
\section{Method}\label{Sec:Method}
To visualize the 3D simulation and to easily compare with the 2D projected observed images, we created a 2D map from the 3D particle distribution for the simulated galaxies by assigning the particles in pixels of 120 by 120 pc, which roughly corresponds to the physical resolution of the simulations, while observing the simulated galaxies face-on. 

\subsection{Radial analysis}
We are interested in the general trend varying with distance from the galaxy's center. Hence, we divide the galaxy into concentric rings based on the galaxy's inclination and position angle up to a certain radius. We decide to separate the galaxies into rings with the largest ring having a radius corresponding to the commonly used $\rm R_{\rm 25}$ to compare the galaxies in rings with relatively similar sizes. This is the radius in which the average $B$-band surface brightness equals 25 $\rm mag\, \rm arcsec^{-2}$. This radius also overlaps with the S/N threshold for our observational constraints. For the $\rm R_{\rm 25}$ of our observations, we take the values found in the literature, which are summarized in Table~\ref{tab:Rings}. Each ring has a width of $0.1\, R_{\rm 25}$, which is still larger than the spatial resolution of our convolved images. We cover each galaxy up to $1\,R_{\rm 25}$. In each ring, we calculate the average surface mass density for the stars $\Sigma_{*}$, the total gas $\Sigma_\mathrm{gas}$, the atomic hydrogen gas $\Sigma_\mathrm{HI}$, the molecular hydrogen gas $\Sigma_\mathrm{H_{2}}$, and the dust $\Sigma_\mathrm{dust}$, together with the average molecular or cold gas fraction $f_\mathrm{mol}$. We also compare the inferred quantities as a function of the oxygen abundance $12+\log\rm (O/H)$, a tracer for the metallicity $Z$, using radial abundance gradients found in the literature (see Table\,\ref{tab:Prop}). We convert the oxygen abundance to Z using the following equation, which assumes the same compositional mixture as the solar neighborhood and a solar metallicity of Z$_\odot$=0.0134:
\begin{equation}
    \mathrm{Z} = 27.4\times10^{\log(\mathrm{O/H})}.
\end{equation}

\subsection{Simulation parameters and initial conditions}
This study is based on a set of isolated galaxy simulations, where we vary the normalized accretion timescale, $\tau_{0}$, of Eq. \ref{eq:accretion} and the $\alpha$ parameter, which governs the mass fraction of the dense clouds of gas particles. Specifically, we vary $\tau_{0}$ from 160\,Myr to 800\,Myr to cover the wide spread of proposed accretion times in the literature \citep[e.g.,][]{Matts12, Asan13, Zhub14, Gallia18, DeVis17,Looze20}. As mentioned in Sect. \ref{Sec:Osaka_GAS}, we vary $\alpha$ from 0.012 to 0.12.
We use the initial-condition setup of an isolated galaxy simulation \citep[see the Appendix of][]{Matsu24} for an NGC 628-like galaxy. In the initial conditions, the gas and old stellar particles are distributed in the rotating disk, reproducing the radial profiles of gas and stellar mass surface density from our NGC 628 observations. The rotation curve of the disk  in the initial
conditions is consistent with that obtained from observations of  NGC 628 by \citet{Aniyan18}. These simulations start with a gas phase metallicity of $0.0001\,Z_{\odot}$. 

We show the evolution of these physical properties in the different rings for the model with  $\tau_{0}=160\,\mathrm{Myr}$ and $\alpha = 0.12$ in Fig. \ref{fig:SGFRZ_1SD} to highlight a couple of aspects. First, the SFR shows a large scatter over time due to the scheme used, yet remains constant on average after 4 Gyr. Second, the stellar mass surface density $\Sigma_{*}$ increases with time, whereas the gas mass surface density $\Sigma_\mathrm{gas}$ decreases slightly with time because of the conversion from gas to stars. Only Z increases steeply with time.

\begin{figure}[!ht]
    \centering
    \includegraphics[width=0.99\linewidth]{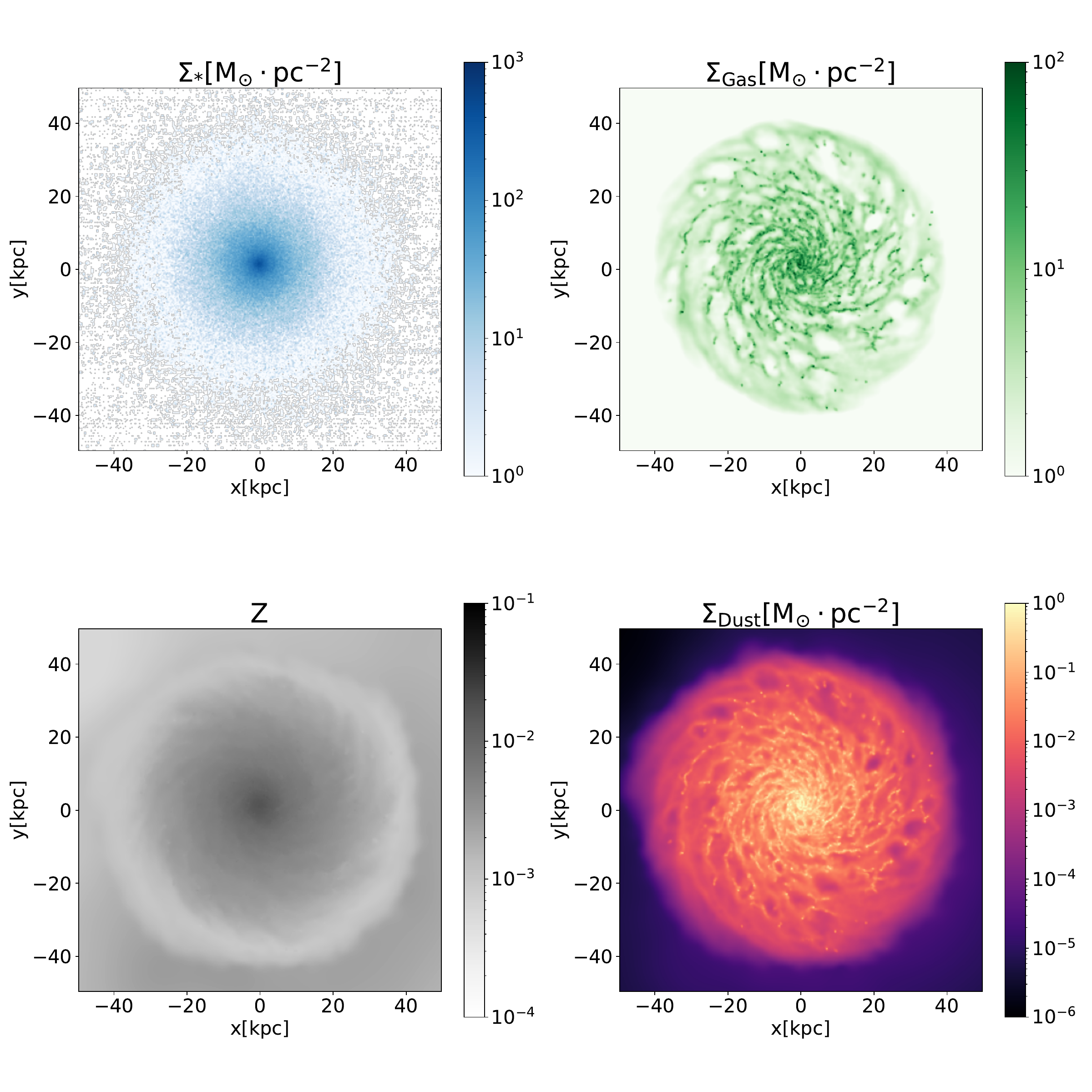}
    \caption{Projected stellar mass surface density $\mathrm{\Sigma_{*}}$ (top left), gas mass surface density $\mathrm{\Sigma_{gas}}$ (top right), gas-phase metallicity Z (bottom left), and dust mass surface density $\mathrm{\Sigma_{dust}}$ (bottom right) for the $\mathrm{\tau_{0} = 160\,Myr}$ galaxy simulated in the \texttt{GADGET4-OSAKA} at snapshot $t=10$ Gyr.}
    \label{fig:StarSim}
\end{figure}

\begin{figure}[!ht]
    %\sidecaption
    %\centering
    \includegraphics[width=\columnwidth]{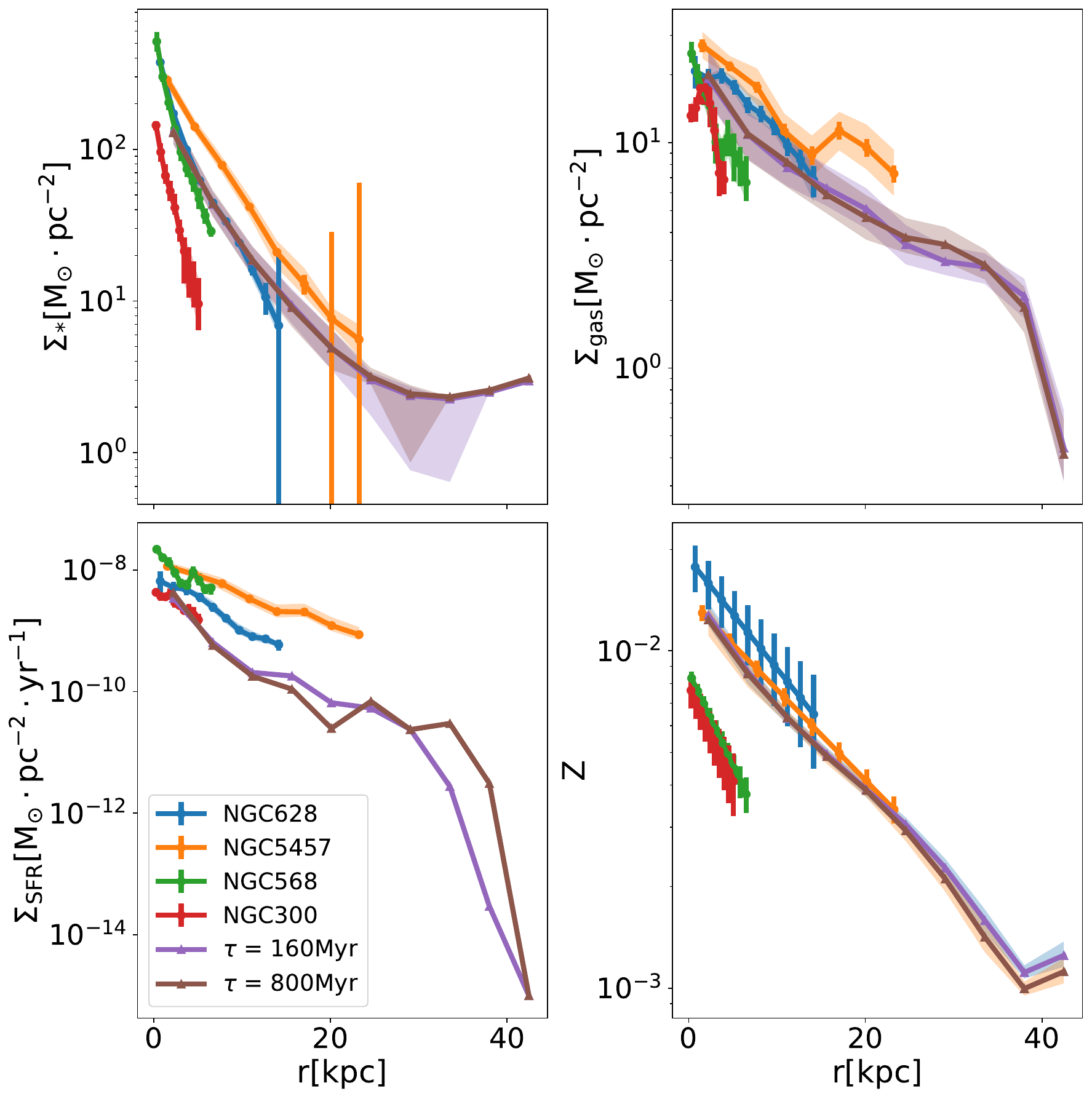}
    \caption{Radial profiles. Here we show $\mathrm{\Sigma_{*}}$ (top left), $\mathrm{\Sigma_{gas}}$ (top right), $\mathrm{\Sigma_{SFR}}$ (bottom left), and Z (bottom right) for NGC\,628, NGC\,5457, NGC\,598, and NGC\,300 (as shown in the legend) compared to the two galaxies simulated in the \texttt{GADGET4-OSAKA} with two different values for the dust-accretion timescales ($\tau_0=160$ and 800 Myr for the purple and brown lines, respectively). The lines indicate the average values and the uncertainties of the rings, and the shaded area is the 1$\mathrm{\sigma}$ within the ring. We note here that $\mathrm{\Sigma_{SFR}}$ shows little dispersion inside the rings in the simulations. The initial conditions for the \texttt{GADGET4-OSAKA} simulations were adjusted to reproduce the observed profiles for NGC\,628 \citep[see the Appendix of][]{Matsu24}.}
    \label{fig:StarSD}
\end{figure}

We show the spatial distribution of these four quantities for the simulated galaxy in Fig. \ref{fig:StarSim} to illustrate how the simulations are able to trace the spiral arms of the galaxy in the gas and dust mass surface density maps, the central concentration of the stellar mass distribution, and the radial Z gradient. To study the variations of the different properties as a function of the galactocentric distance, we derived radial profiles from these maps and compare these with the radial profiles obtained from the observations. The comparison is shown in Fig. \ref{fig:StarSD}. 

For NGC\,628, the observed stellar and gas mass surface density radial profiles show good agreement with simulations, albeit with the Z predicted by the simulations being in the low range of the observations. While the central SFR surface density for NGC\,628 is reproduced by the simulations, the radial decline of the observations is shallower than the simulated trend. We note that SFR surface densities in the simulations with different accretion timescales differ in the outskirts. This is because stochastic star formation is more pronounced in regions with low star formation rates rather than dust accretion affecting star formation. The other galaxies in our sample are added here for comparison. NGC\,300 and  NGC\,598 show lower Z and slightly lower gas masses than NGC\,628 and NGC\,5457, and in particular, NGC\,300 differs in the overall values of stellar mass and SFR surface density from the other galaxies in the sample. 

The models with $\tau_{0}=160\,\mathrm{Myr}$ and $\alpha = 0.12$ represent our fiducial model as \citet{Roma22a,Roma22b} used these parameters for their model results. By varying these two parameters, we investigated the effect of $\tau_{0}$ and $\alpha$ variations on the simulations to deduce which combination best reproduces the observations.
This provides greater insight into the dominant dust formation and destruction processes at work in NGC\,628 and the other galaxies. We compare the radial profiles of the dust mass surface density, the DGR, the DZR, and the SLR.

%------------------------------------------------------------
%\newpage 
\section{Results}\label{Sec:Results}
\subsection{Dust and dust-scaling quantities}\label{Sec:Results_DustScaling}
\begin{figure}[!ht]
    %\sidecaption
    \centering
    \includegraphics[width=\columnwidth]{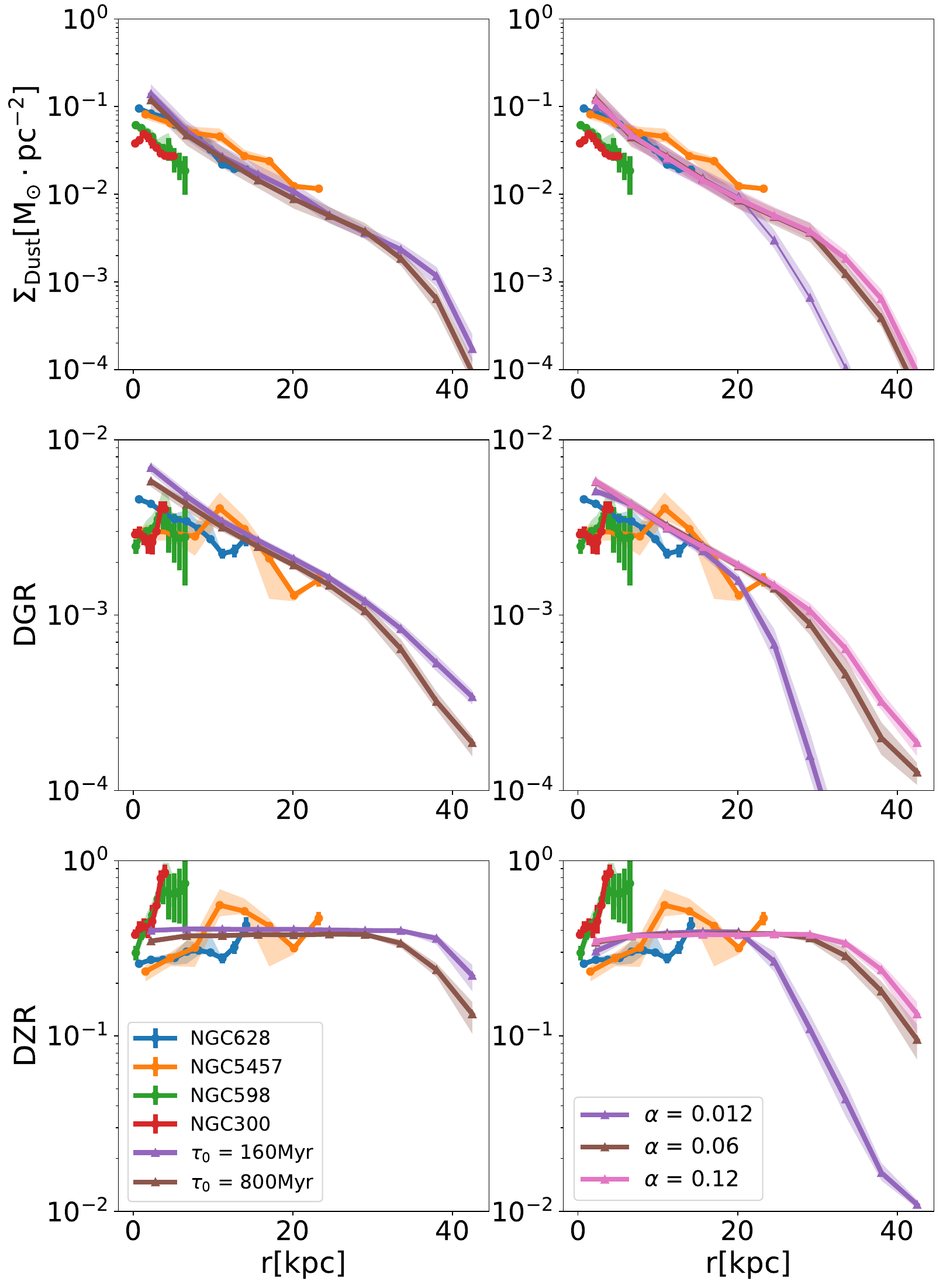}
    \caption{$\mathrm{\Sigma_{dust}}$, DGR, and DZR ratio versus physical radii for observations and simulations varying with accretion timescale (left) and $\alpha$ (right). The lines are shown in the same colors as in Fig.\ \ref{fig:StarSD}.}
    \label{fig:Dust_tau}
\end{figure}

The three panels of Fig. \ref{fig:Dust_tau} show --- from top to bottom --- $\mathrm{\Sigma_{dust}}$, DGR, and DZR for the four observed galaxies and the set of simulated galaxies. We can divide the observed galaxies into two pairs:  NGC\,628 and NGC\,5457, and NGC\,598 and NGC\,300. NGC\,628 and NGC\,5457 have higher $\mathrm{\Sigma_{dust}}$, higher DGR, but lower DZR relative to NGC\,598 and NGC\,300. The difference in the DZR indicates the potential high efficiency of dust accretion in NGC\,598 and NGC\,300.

The top-left panel of Fig. \ref{fig:Dust_tau} shows that the simulated dust mass surface densities are in relatively good agreement with the observed values for NGC\,628 after correcting for element-dependent accretion (see Sect. \ref{Sec:Data_Sim_Dust}). In addition, dust mass surface densities do not vary significantly when changing the dust-accretion timescale. We can only see the effect of the various accretion timescales in the outskirts, where Z is the lowest (see the bottom-right panel in Fig. \ref{fig:StarSD}). The same trend continues when comparing the DGR and the DZR, with only minor differences in the outskirts. We notice, especially in the DZR, that the ratio converges to 0.4, which means that the uncorrected DZR converges to 1 for all rings within 30\,kpc. Beyond 30\,kpc, corresponding to a Z of $2\times10^{-3}$, the DZR is not at the maximum value. The outskirts exhibit a lower Z, which means that accretion is slower than in the higher-Z rings. The slow accretion rates in the outskirts could explain why the difference in dust mass between the simulations is a factor of 2.5 or less. We show how we verify this below, as an increase in accretion rates should raise the abundance of small dust grains.

In the right column of Fig. \ref{fig:Dust_tau}, we show the effect of varying $\alpha$. We note that the radial gradients change more in the simulations varying in $\alpha$ than in those varying in $\tau_{0}$. We see that $\mathrm{\Sigma_{dust}}$ remains unchanged for the simulations within the first 20\,kpc, but beyond this radius, $\mathrm{\Sigma_{dust}}$ drops steeply with the slope exhibiting an anti-correlation with $\alpha$. These drops with $\alpha$ also appear in the DGR and DZR, which we expect, as only the dust content varies between simulations. Lower values of $\alpha$ increase the domination of the diffuse ISM, which would imply that dust has more difficulty growing in the outskirts where Z and dense gas fraction also drop, hinting at the importance of the  build up of dust in the dense ISM. 

In the end, we are not able to constrain the most appropriate  $\tau_0$ value, as accretion appears to saturate quickly, using up all the available metal. Also, varying the relative contributions of the diffuse and dense ISM phases seems to have a negligible effect. Therefore, it remains unclear as to what causes the different DZRs in the four galaxies. Apart from the parameters that we studied, the average density of the dense ISM could differ among this set of galaxies or dust composition could differ from one galaxy to another (which we currently do not consider). For instance, if the dust is more emissive in these galaxies than the typical MW dust, this would mean that we have overestimated dust masses. The overestimation could be the case, as the DZRs are too high to be realistic. 

\subsection{Small-to-large grain abundance ratio}

\begin{figure}[!ht]
    \centering
    \includegraphics[width=0.99\linewidth]{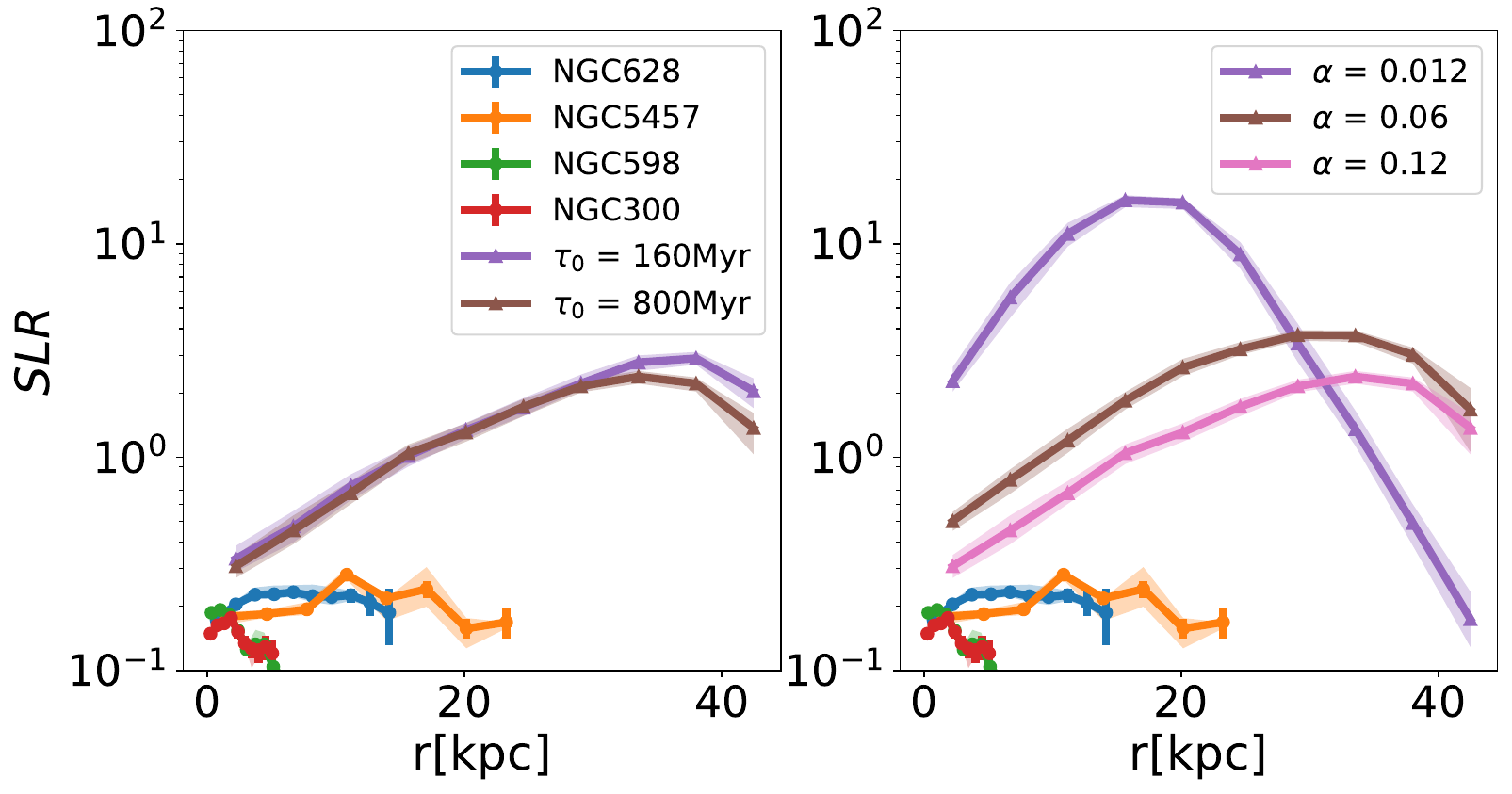}
    \caption{SLR versus physical radius for observations and simulations varying with accretion timescale (left) and $\alpha$ (right). The lines are shown in the same colors as in Fig.\ \ref{fig:StarSD}.}
    \label{fig:DustSLR_tau}
\end{figure}
Both panels of Fig. \ref{fig:DustSLR_tau} show the SLRs for the observations. We note here that we can separate the galaxies into the same pairs: NGC\,628 and NGC\,5457 have higher SLRs than NGC\,598 and NGC\,300. The SLRs of the former group remain constant with radius, whereas those of the latter decrease. In this figure, we show the SLR comparison between the observations and the simulations at 10\,Gyr. In general, the SLRs in the simulations increase with radius. As we do not see any variation in SLR in the inner 30\,kpc between simulations, we can conclude that the variation in the accretion timescale does not affect the grain-size distribution. The increase in the SLR with radius might be affected by the radial decrease in the dense gas fraction, which means that dust evolution mechanisms in the diffuse ISM become more efficient relative to those in the dense ISM as radial distance increases. This variation implies that coagulation may become less efficient and shattering more efficient at larger galactocentric distances. We also see the expected higher SLR beyond 30\,kpc in the simulation with a shorter accretion timescale. This higher SLR confirms that accretion variations are noticeable in low-Z conditions. 
The DZR and SLR decrease beyond 35\,kpc, which suggests that accretion becomes inefficient if Z is too low.

The right panel of Fig. \ref{fig:DustSLR_tau} shows the effect of $\alpha$ in the SLR. The radii at which the SLR varies overlap with the radii we observe variation in the $\Sigma_{dust}$, DGR, and DZR from the right side of Fig. \ref{fig:Dust_tau}. The simulations with the lowest $\alpha$, in which the diffuse ISM is the most dominant, have the highest SLRs. This result is in line with less efficient coagulation and more efficient shattering. We also note that the turnover point from increasing to decreasing SLR happens closer to the center of the  galaxy in the simulations with lower $\alpha$. We observe variations in the DZR at the same radius. As coagulation requires a high gas density, the drop-off in DZR and SLR at outer radii should be caused by the dominance of stellar dust production, as not all metals get accreted onto dust grains.

We note that the SLR increases for NGC\,628 up to 5\,kpc and slowly decreases beyond that, whereas the ratio constantly decreases with radius for NGC\,598 and NGC\,300. This observed trend should imply that the simulations with lower $\alpha$ values represents NGC\,628 the best, but that would produce, in turn, excessively high SLRs, which indicates that the simulations are too efficient in producing small grains. This may appear to be contradictory with the underestimation of the simulated PAH fraction discussed in \citet{Matsu24}, but we specify that the PAH fraction is only a fraction of the small grains. The coagulation of PAH grains also does not directly result in grains large enough to be considered part of the large grains.  In Sect. \ref{Sec:Disq_DensDiff}, we discuss the possible causes of the overproduction of small grains or the underproduction of large grains and how we can improve the simulations in the future.
%------------------------------------------------------------
\section{Discussion}\label{Sec:Disq}

\subsection{Accretion timescale}\label{Sec:Disq_Accretion}
Our results suggest that accretion might not explain the differences in the radial profiles of dust abundance between observations and simulations.
Accretion is a mechanism proposed to solve the missing dust budget problem observed in the ISM  \citep[e.g.,][]{Dwek98, Hira99}, as stellar dust sources are not sufficient to account for the dust observed in the ISM. Different dust-accretion prescriptions have been proposed in the literature: \citet{Matts12} formulated the accretion timescale as dependent on $\mathrm{\Sigma_{SFR}}$, $\mathrm{\Sigma_{gas}}$, $Z$, the DZR, and a free parameter defined as efficiency. With this prescription, subsequent studies predicted a wide range of accretion time scales: $\sim 20-200$\,Myr was obtained by \citet{Matts12b} to describe observations of nearby galaxies without Herschel observations, values of 50\,Myr-1\,Gyr \citep{Gall21} and 200\,Myr--1\,Gyr \citep{DeVis17} were obtained from chemical and dust evolution models applied to observations, and \citet{Looze20} derived timescales of around 400\,Myr with chemical evolution models using customized star formation histories to describe a large galaxy sample with a wide range of physical properties. 

The simulations presented here make use of a parametrization of the accretion timescale (see Eq.2) that takes into account the size of the dust grains where the accretion of gas-phase metals onto the surface of the grains takes place \citep{Asan13} and the sticking efficiencies of the process \citep{Zhuko16}. The parametrization is applied to each SPH particle with different gas density, temperature, and Z. For the normalization of the accretion timescale in Eq.2 ($\tau_{0}$) we have chosen 160\,Myr and 800\,Myr, which are values within the range proposed by previous studies.

\begin{figure*}[!ht]
    \centering
    \includegraphics[width=\linewidth]{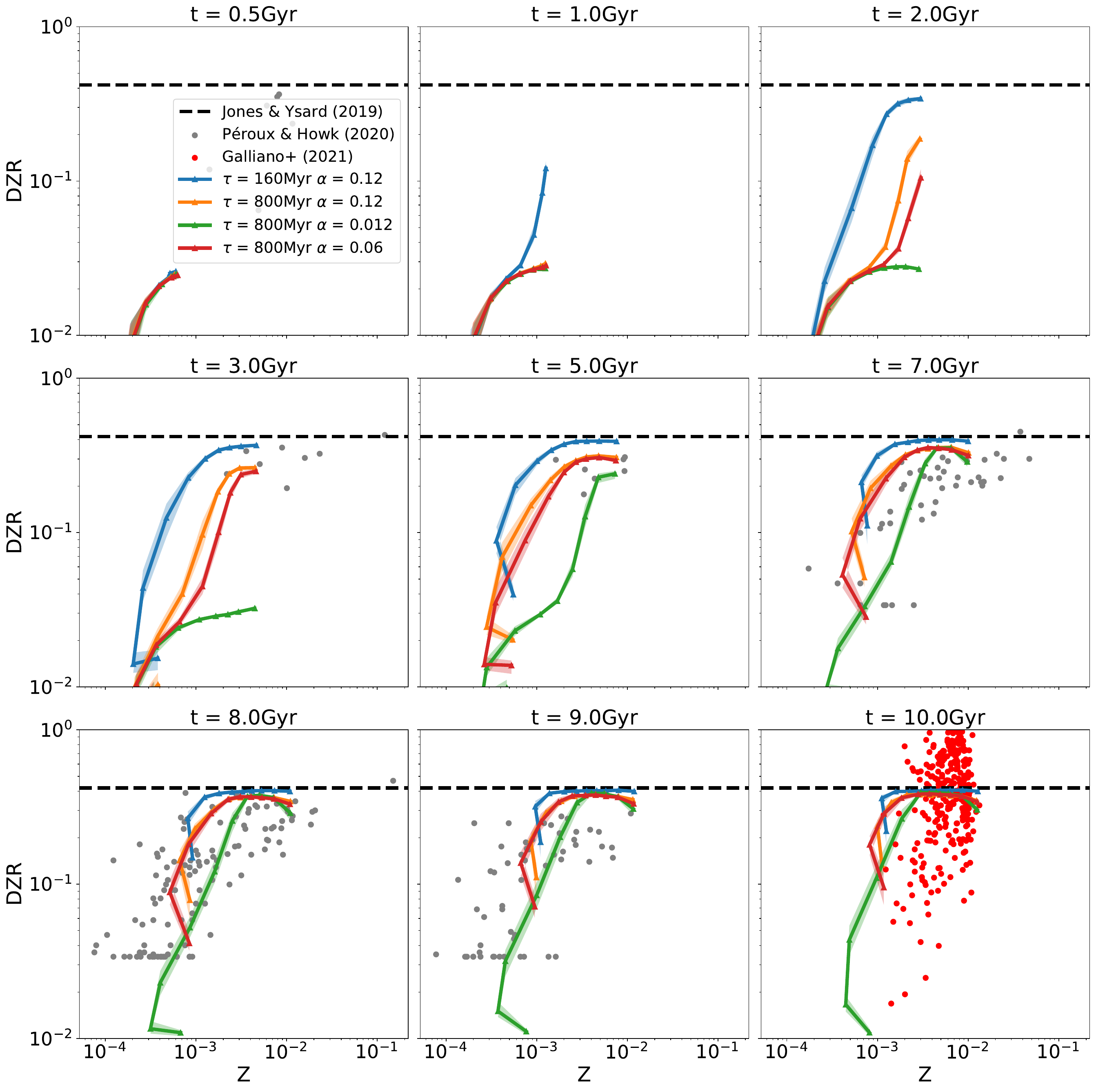}
    \caption{DZR versus Z using the averages in the radial annuli for our simulated galaxy at the different snapshots. We compare the snapshot at t=10\,Gyr with observations of the nearby Universe.}
    \label{fig:DZR_time}
\end{figure*}

The results shown here indicate that varying the accretion timescale has a negligible effect on the dust masses in observed galaxies except in their outskirts. We show this in Fig. \ref{fig:DZR_time}, where we plot the average DZR varying with the average Z separated in the radial annuli used in the earlier comparisons for each snapshot. We compare the snapshots with global properties for galaxies at different redshifts \citep{Peroux20, Gall21} assuming that the 10\,Gyr snapshots of our simulated galaxies correspond to the local Universe. We see in the first snapshots that the shorter accretion timescale does allow for faster dust build up, as the DZR is higher for the simulation with the shortest timescale. The simulation with the longer accretion timescale catches up at $Z>0.003$ once the DZR converges to the maximum value.
As explained in Sect. \ref{Sec:Results_DustScaling}, the DZR in the simulations converges to its maximum value, and then accretion stops. Figure \ref{fig:DZR_time} shows that the accretion timescales chosen here may be too short, pointing to longer times, which would be at the high end of the ranges proposed by previous studies \citep{DeVis17,Gall21,Looze20}. Moreover, the efficiency of accretion is degenerate with that of SN destruction \citep[e.g.,][]{Calura23}, and therefore our comparison does not allow us to robustly restrict the accretion timescale.

Our comparison with observations suggests that the general trend of the observed DZR is the convergence to its maximum value if we correct the resulting dust masses for the differential accretion of different elements (see Sect. \ref{Sec:Data_Sim_Dust}). This correction suggests that we might have to add the elemental constraints and thereby lower the accretion rate (e.g., increase the accretion timescale) as a result. Oxygen for example is more abundant than carbon in the ISM \citep[e.g.,][]{Jones19}, but carbon gets depleted much easier than oxygen depending on Z \citep{Konsta24}, which suggests that grain growth depends on the chemical composition of grains.

We notice that accretion dominates the dust build up above a critical Z, as shown in the outskirts of the simulated galaxies in Fig. \ref{fig:Dust_tau}. Therefore, we have to study accretion timescales in pristine low-Z environments and decipher what dictates the transition. Our current set of large spiral galaxies do not reach these low metallicities. In the future, it would be interesting to calibrate the accretion timescale in the simulation to low-$Z$ dwarf galaxies such as the
Magellanic Clouds %%LMC and SMC or
and the sample of the Dwarf Galaxy Survey \citep[][with metallicities as low as 1/50$\mathrm{Z_{\odot}}$]{Madd13}, where the metallicities might be low enough and the resolution large enough to observe these pristine environments, which are similar to the outskirts of our simulations.

\subsection{Diffuse and dense ISM treatment}\label{Sec:Disq_DensDiff}
Our results suggest that the differences in dust evolution between our observations and simulations relate to differences in the grain size distribution. Some major processes that alter grain sizes depend on the ambient physical conditions of the ISM.
Taking into account the dense gas properties in galaxy simulations with limited spatial resolution remains difficult. This is why simulation studies such as the present one apply subgrid models to constrain the fraction of hydrogen locked up in the dense phase and assume temperatures and densities. However, we need a correct treatment of the diffuse and dense ISM because the dust evolution mechanisms vary differently within these two phases. 

Observers use the ratio of molecular hydrogen gas mass to total gas mass as a tracer of the mass fraction of the dense gas. Figure \ref{fig:MH2_alpha} shows the comparison of $\Sigma_\mathrm{H_{2}}$ between observations assuming a Z-independent CO-to-$\mathrm{\Sigma_{H_{2}}}$ conversion factor $X_\mathrm{CO}$ for the Milky Way \citep{Bola13, Sand13} and our simulations. Despite the agreement in $\Sigma_\mathrm{gas}$, $\Sigma_\mathrm{H_{2}}$ is underestimated in the simulations compared to NGC\,628 and NGC\,5457. We tried other methodologies, such as that of \citet{Hunt15}, and we recover this same result, regardless of the chosen ${X_\mathrm{CO}}$. This underestimation in molecular gas mass influences the relevant processes in the dense ISM (coagulation in particular), which would lower the conversion from smaller grains to larger grains. The increased weight for the diffuse ISM phase could overestimate the importance of shattering in the simulations, enhancing the conversion from large to small grains. We expect a higher SLR in the simulations compared with observations based on this difference, exactly what we observe in Fig. \ref{fig:DustSLR_tau}. However, the difference in molecular gas content might not explain the difference. We note that the simulated molecular gas mass profiles agree with NGC\,598 and NGC\,300, but Fig. \ref{fig:DustSLR_tau} shows that NGC\,598 and NGC\,300 have lower SLR than NGC\,628 and NGC\,5457 and therefore diverge more from the simulated SLR. The results for NGC\,598 and NGC\,300 are at odds with our expectation that a lower dense gas fraction would result in shattering becoming more important in the diffuse ISM at the expense of coagulation in the dense ISM, which should result in a higher SLR. The fact that the SLR is lower in NGC\,598 and NGC\,300 suggests that the different physical properties and formation histories might have impacted the SLR, which makes the comparison with the simulations tailored to NGC\,628 far from straightforward.

The biggest differences between our observations and the simulations in our results occur in the SLR, implying that the simulation subgrid models for the diffuse and dense ISM are not able to fully capture the intricacies of the multi-phase gas distribution. We show in Figs. \ref{fig:StarSD} and \ref{fig:Dust_tau} that the accretion timescale only has a minor effect on $\Sigma_\mathrm{dust}$ in the outer rings, where physical conditions cause accretion to be slower. We therefore deduce that the biggest driver of the difference in the SLR between observations and simulations is the difference in the balance of the size-altering processes ---grain shattering and coagulation--- between the two. We did experience problems in tracing the dense and diffuse ISM  when modeling $\mathrm{H_{2}}$ formation, as the simulated $\mathrm{\Sigma_{H_{2}}}$ radial profile varies with $\alpha$ as shown in Fig. \ref{fig:MH2_alpha}. We do expect $\mathrm{\Sigma_{H_{2}}}$ to differ when we change the allowed dense gas fraction, but we only noted a strong deviation in $\mathrm{\Sigma_{H_{2}}}$  when we simulated our isolated galaxy using an $\alpha$ value of 0.012. This strong variation either indicates that we cannot use $H_{2}$ as a dense gas tracer or that these dense gas fractions are of-limit to reproduce the observations. Therefore, we have to first improve the agreement in the H$_2$ mass fractions between the observations and simulations before changing the efficiencies of specific dust-evolution processes. 

A potential dust-related problem in the simulations is the diffusion treatment incorporated in the simulations. Diffusion transfers the dust grains from the dense to the diffuse ISM, which can benefit the formation of small grains. The subgrid model using the relative velocities of the particles to approximate the shear and therefore the transfer between particles. Diffusion can be too efficient and can artificially lower the  masses of the large dust grains and enhance the formation of  small grains through shattering. \citet{Roma22a} already obtained a high SLR using simulations without using diffusion or by reducing their diffusion efficiency by a factor of 0.1. This implies that adjusting the parameters involving diffusion will not result in a significant reduction of the simulated SLR, which would otherwise allow agreement with the observed ratios of our galaxy sample.

However, dust also transfers from the dense to the diffuse phase whenever the dense gas cloud is at the end of its life cycle and dissipates. This process may destroy part of the dust due to the increased turbulence during the dissipation before it becomes part of the diffuse ISM (A. Jones et al. in prep). This process happens on the subgrid model level and requires testing on smaller scales than those provided by the simulations presented in this work. 

Dust growth in our simulations depends strongly on the dense gas fraction. Future simulations need careful calibrations for the different phases and the treatment of the different dust evolution processes in order to obtain the low SLRs seen in our observations. This calibration is the main priority for future works.

\subsection{Comparison with other galaxy simulations}\label{Sec:Disq_Sim}
When we compare our results with simulations in the literature, we note that some studies succeed in reproducing Milky Way-type dust properties. \citet{Roma22a} use the same simulation setup as that adopted in the present paper, but calibrate the single galaxy simulation to the Milky Way. They compare the results of their simulations with the observations presented in \citet{Rela18}. 

The SLR values obtained by \citet{Roma22a} from their simulation were only higher by 0.2 dex than the observed values given in \citet{Rela18}, and showed similar radial variations to those seen in observations. Therefore the calibration to galaxies other than the Milky Way could cause this agreement to disappear. The calibration to $\mathrm{\Sigma_{*}}$, $\mathrm{\Sigma_{SFR}}$, and Z does not seem to produce a unique galaxy that matches well with the observations. The fine-tuning of the simulations to fit observed galaxies therefore requires more in-depth investigations in future work in order to also take the $H_{\text{2}}$ molecular gas fraction into account.

\citet{Gran21} implemented subgrid models for the dust with the two-size approximation \citep{Hira15}. The simulations use GADGET-3 with a locally developed subgrid model MUPPI \citep{Vale19}, which includes treatment for molecular gas and implements the same dust evolution processes as those discussed in the present paper. MUPPI calculates the dense ISM fraction based on the cold gas fraction constraints from observations. The dust evolution processes also make use of this detailed ISM phase grid. \citet{Gran21} matched results in the DGR and the small-to-large grain ratio with observational results from \citet{Rela20} using their set of low-spatial-resolution and high-spatial-resolution galaxies. The $\Sigma_\mathrm{dust}$ is overestimated by 0.5 dex in their low-resolution simulations, whereas the overestimation is reduced to 0.2 dex (within the uncertainties of observations) in their high-resolution simulations. \citet{Parent22} provide a similar example of this phenomenon; these authors used the \citet{Gran21} models, yet they only constrained the agreement in the dust properties for Milky Way-type galaxies.

A more plausible point of contention is the dynamical evolution of galaxies compared to the Milky Way. Several studies show that the Milky Way has a relatively uneventful merger history \citep[e.g.,][]{Hamm07}, with long periods of evolution without any interaction. This long non-interactive history is reminiscent of our single-galaxy simulations. It could be that the nearby galaxies in our sample experienced interactions with other objects at later times, which could have caused additional inflows, starbursts, and/or dynamical mixing, and would violate the assumptions we make when we apply single isolated galaxy simulations. The addition of these phenomena will impact the interaction of dust grains, potentially enhancing grain--grain interactions or astration. It is therefore important to constrain at least the star formation history and inflow gas history and explore how differences in the star formation histories of the simulated galaxies influence comparisons with observations.

\subsection{Improvements}
We are currently working with state-of-the-art single-galaxy simulations, incorporating dust evolution with a full treatment of the grain-size distribution. The simulations are able to reproduce spatially resolved features in the stellar, total gas, and Z components of NGC\,628. However, the mismatches in the SLR and the molecular-gas-mass surface density may be signs that the current set of simulations require improvement. In this section, we highlight some possible pathways for such improvements.

The focus could be put on applying detailed subgrid models that allow a more detailed separation of the different ISM phases than the parametrization used here in terms of the $\alpha$ parameter. This would have consequences for the global efficiency of the different dust-evolution mechanisms and therefore for the total dust mass surface densities and SLRs.  

Another improvement could be made in terms of the coupling of dust and chemistry. Dust chemistry is run in the simulations separately from the rest of the chemistry. A further step would be to combine this coupling by adding the key element approximation \citet{Zhu08} in the dust evolution scenario \citep[e.g.,][]{Choban22, Choban24}. The key element approximation will be incorporated in GADGET simulations in a future work (K. Matsumoto et al. in prep.).

Third, a parametrization of the dense ISM \citep{Henne24} that is  more detailed than the single combination of density and temperature assumed by the present simulations could allow a more detailed representation of the different ISM phases. 
As we show in Sect. \ref{Sec:Results_DustScaling}, the SLR is sensitive to the treatment of the dense gas fraction of the subgrid models. It may be useful to explore whether or not a different combination of density and temperature that separates the dense and diffuse ISM can lead to better agreement in the comparison of the molecular gas masses and the SLR. 

Finally, we could focus on the effect of the star formation history on the efficiency of the dust-evolution mechanisms. We calibrated the single-galaxy simulation to agree with the stellar mass and total gas mass surface density, Z, and present-day SFR, assuming a constant SFR throughout its evolution. The simulations show negligible growth in stellar mass and gas mass surface density (see Appendix \ref{App:Time}). If a galaxy creates the bulk of the stars and the metals at later times, we could observe a galaxy at a different stage of dust accretion. For instance, NGC\,598 and NGC\,300 could be at an earlier stage in their  evolution than NGC\,628 and NGC\,5457, as their metallicities are on average lower. Hence, NGC\,598 and NGC\,300 might be in the dust-build-up phase where dust growth has not converged, which could explain the lower SLR observed for these galaxies. Their metallicities are rather above $2\times10^{-3}$, which is when accretion should play a role, and their DZR is higher and more likely converging, contradicting this claim.

\section{Conclusion}\label{Sec:Conq}
In this study, we calibrated a single galaxy simulation to agree with the nearby galaxy NGC\,628 in terms of stellar mass surface density, gas mass surface density, and Z. With the aim of studying how dust forms and evolves in NGC\,628, we compared the observed dust and metal scaling relations and the SLRs to values obtained in a simulated isolated galaxy with different dust accretion timescales and dense phase ISM fractions. The dense gas mass fraction in the ISM is incorporated in the simulations via the parameter $\alpha$, which regulates the fraction of the gas mass in the form of dense gas in each simulated particle. 

We find that the simulated dust mass surface density always agrees with that observed for NGC\,628 once we lower the dust mass surface densities based on the available metals that can be locked up in dust grains.
The DZR converges to its maximum value for the inner disk with metallicities above $2\times10^{-3}$. We observe differences in the dust mass for our simulations with different accretion timescales in rings with lower metallicities, with the shorter timescales corresponding to higher dust masses. This coincides with a non-converging DZR. 
In the future, we need to add the key-element approximation to allow a natural constraint on the metal budget rather than implementing an arbitrary number. Furthermore, we need to target low-Z galaxies (Z < $2\times10^{-3}$) to constrain the timescales of metal accretion onto pre-existing dust grains. These additions will allow us to obtain a more up-to-date view of dust accretion at various metallicities. Once we constrain the accretion timescale at the various metallicities using these simulations and data sets, we will have strong constraints with which to understand the physical process in detail, which will bring us closer to deciphering the origin of interstellar dust.

The SLR is overestimated in all simulations presented here by at least a factor of 2. We have not been able to find a combination of accretion timescales and $\alpha$ values (mimicking different dense gas mass fractions) that reproduces the values derived from observations. The SLR ratio increases toward the outer part of the galaxy, as seen in observations of NGC\,628 and NGC\,5457. However, the increase in the simulated SLR with distance from the galaxy center is steeper than that seen in these observations. The radially increasing trend could be related to the increased diffuse ISM fraction toward the outer parts of the galaxy. An increase in the diffuse gas mass fraction would enhance the shattering of large grains into small ones, and would therefore result in an increase in the SLR. At even larger radii, stellar dust production and the injection of large dust grains into the ISM might overtake shattering, causing the SLR to decrease. We find significant variation in the radial trend of the SLR for the lowest $\alpha$ value compared to $\alpha$=0.12 and $\alpha$=0.06. In this case, the SLR peaks closer to the center of the galaxy. For such low $\alpha$ values, the gas mass fraction in the form of diffuse ISM is significantly larger than for higher $\alpha$ values, which enhances the efficiency of shattering and causes an increase in the SLR at low galactocentric distances. These results highlight the need to apply a robust subgrid model that accurately reproduces the different ISM phases in order to find agreement with the relative abundances of small and large dust grains derived from observations.\\

\begin{acknowledgements}
We would like to thank the referee for the useful comments which improved the clarity of the paper. S.v.d.G, I.D.L. and M.P. have received funding from the European Research Council (ERC) under the European Union's Horizon 2020 research and innovation programme DustOrigin (ERC-2019-StG-851622), from the Bijzonder Onderzoeksfond (BOF) through the starting grant (BOF/STA/202002/006) and from the Flemish Fund for Scientific Research (FWO-Vlaanderen) through the research project G023821. KM is supported by FWO-Vlaanderen through the grant number 1169822N. MR acknowledges the support from the project PID2020-114414GB-100, financed by MCIN/AEI/10.13039/501100011033. 
Some of the numerical simulations used in this study were carried out at XC50 at the National Astronomical Observatory of Japan, and {\sc SQUID} at the Cybermedia Centre, Osaka University as part of the HPCI system Research Project (hp230089, hp240141). 
This work is supported in part by the MEXT/JSPS KAKENHI grant numbers 20H00180, 22K21349, 24H00002, and 24H00241 (K.N.). 
K.N. acknowledges the support from the Kavli IPMU, the World Premier Research Centre Initiative (WPI), UTIAS, the University of Tokyo.  

\end{acknowledgements}

\bibliographystyle{aa} % This is the style of the bibliography (aa.cls)
\bibliography{refe}

\appendix

\section{Dust mass maps}\label{App:dustmethod}
\begin{figure}[!ht]
    \centering
    \includegraphics[width=\linewidth, clip=true, trim=40mm 140mm 100mm 80mm]{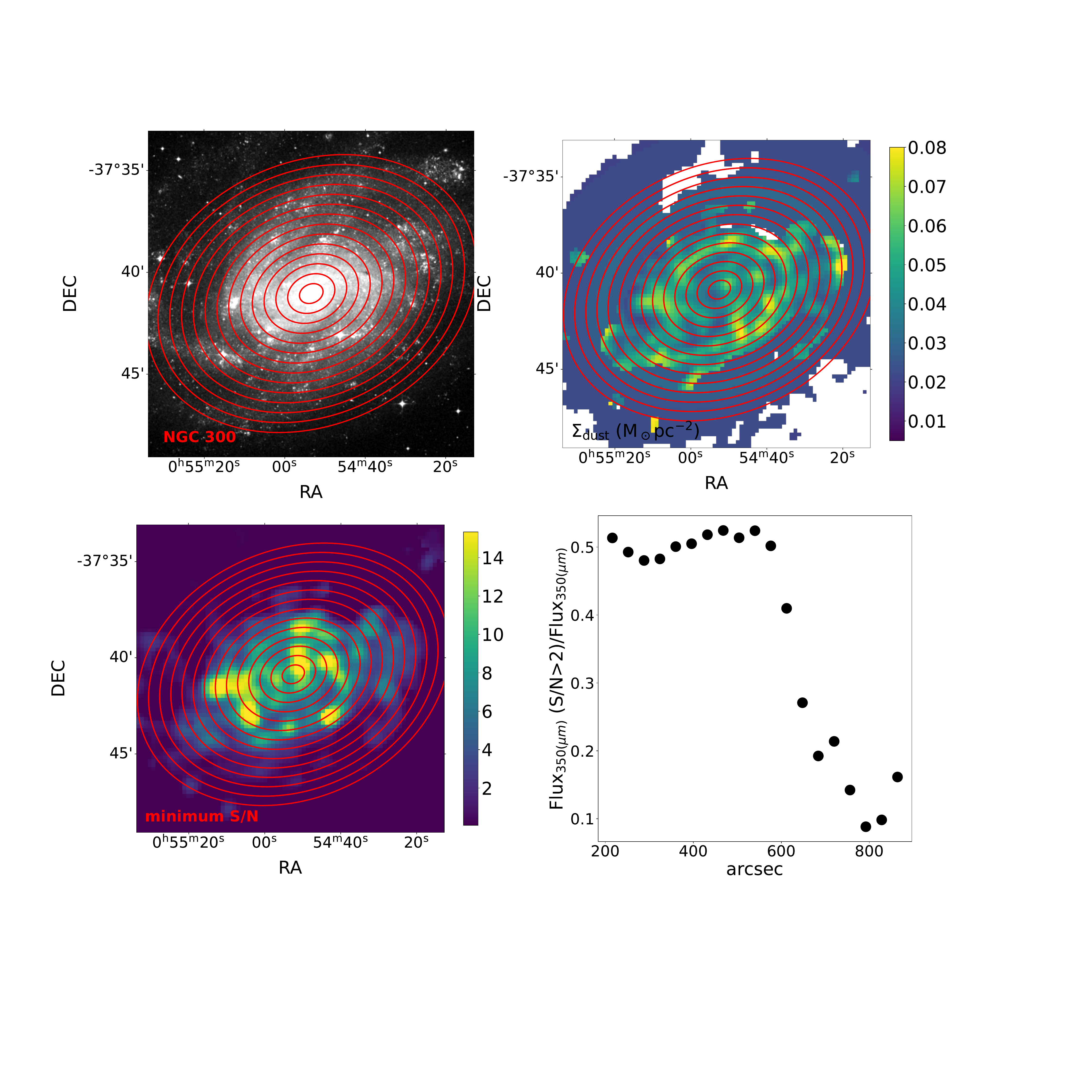}
    \caption{Spatially resolved images of NGC\,300. Top left: B-band image of NGC\,300 with concentric annuli of 36\,arcsec width. Top right: Dust mass surface density of the whole disk. Bottom left: S/N map corresponding to the lowest S/N of all the observed fluxes in each pixel. Bottom right: Total 350$\mathrm{\mu m}$ flux of pixels having an S / N greater than 2 in the 350\,$\mathrm{\mu m}$ band divided by the total 350\,$\mathrm{\mu m}$ flux of the pixels with low S/N in all bands for each ring.}
    \label{fig:dustmethod_300}
\end{figure}

\begin{figure}[!ht]
    \centering
    \includegraphics[width=\linewidth, clip=true, trim=40mm 140mm 100mm 80mm]{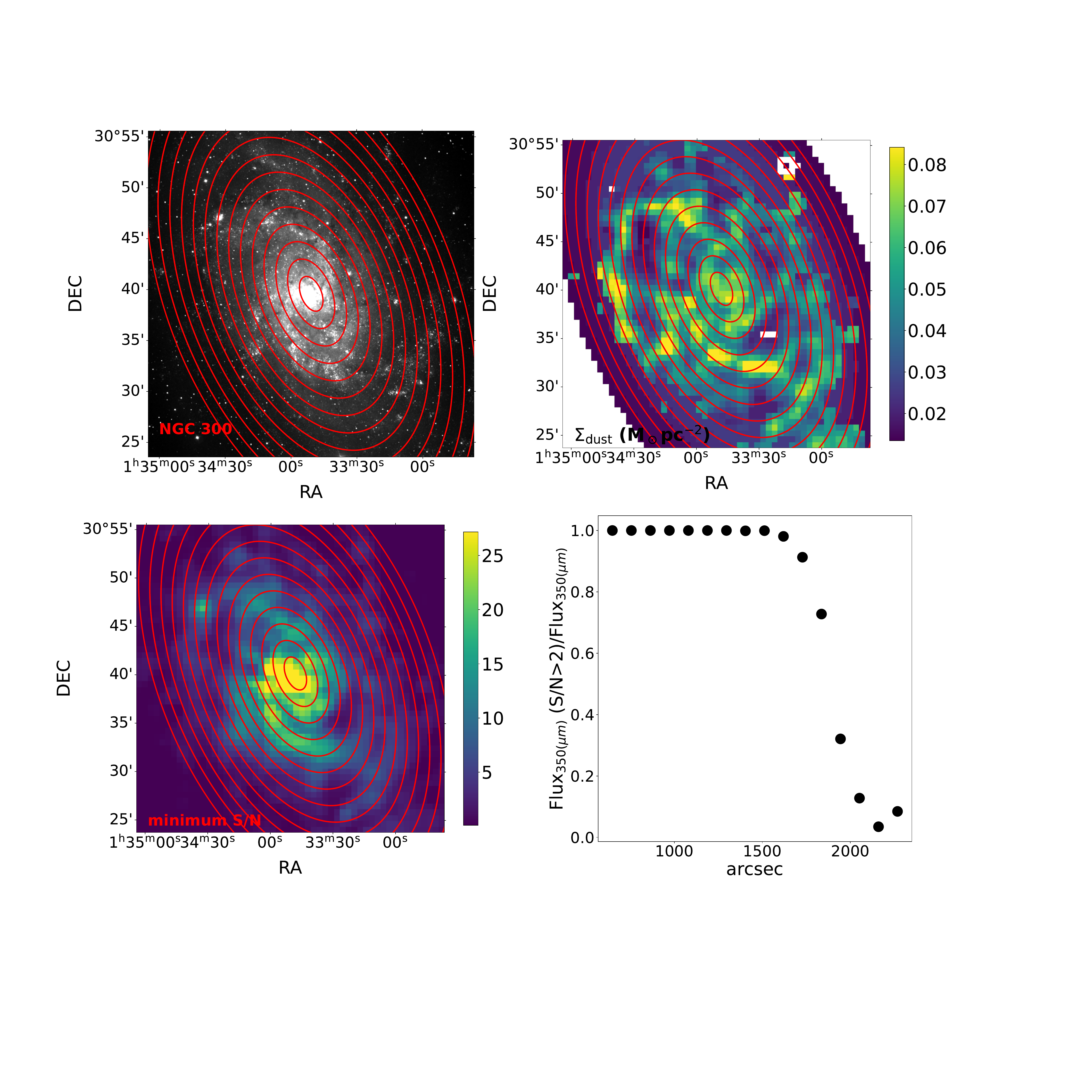}
    \caption{Spatially resolved images of NGC\,598. Top left: SDSS $g$-band image of NGC\,598 with concentric annuli of 108\,arcsec width. Top right: Dust mass surface density of the whole disk. Bottom left: S/N map corresponding to the lowest S/N of all the observed fluxes in each pixel. Bottom right: Total 350$\mathrm{\mu m}$ flux of the pixels having a S/N above 2 in the 350$\mathrm{\mu m}$ band divided by the total 350$\mathrm{\mu m}$ flux of the pixels with low S/N in all bands for each ring.}
    \label{fig:dustmethod_598}
\end{figure}

\begin{figure}[!ht]
    \centering
    \includegraphics[width=\linewidth, clip=true, trim=40mm 140mm 100mm 80mm]{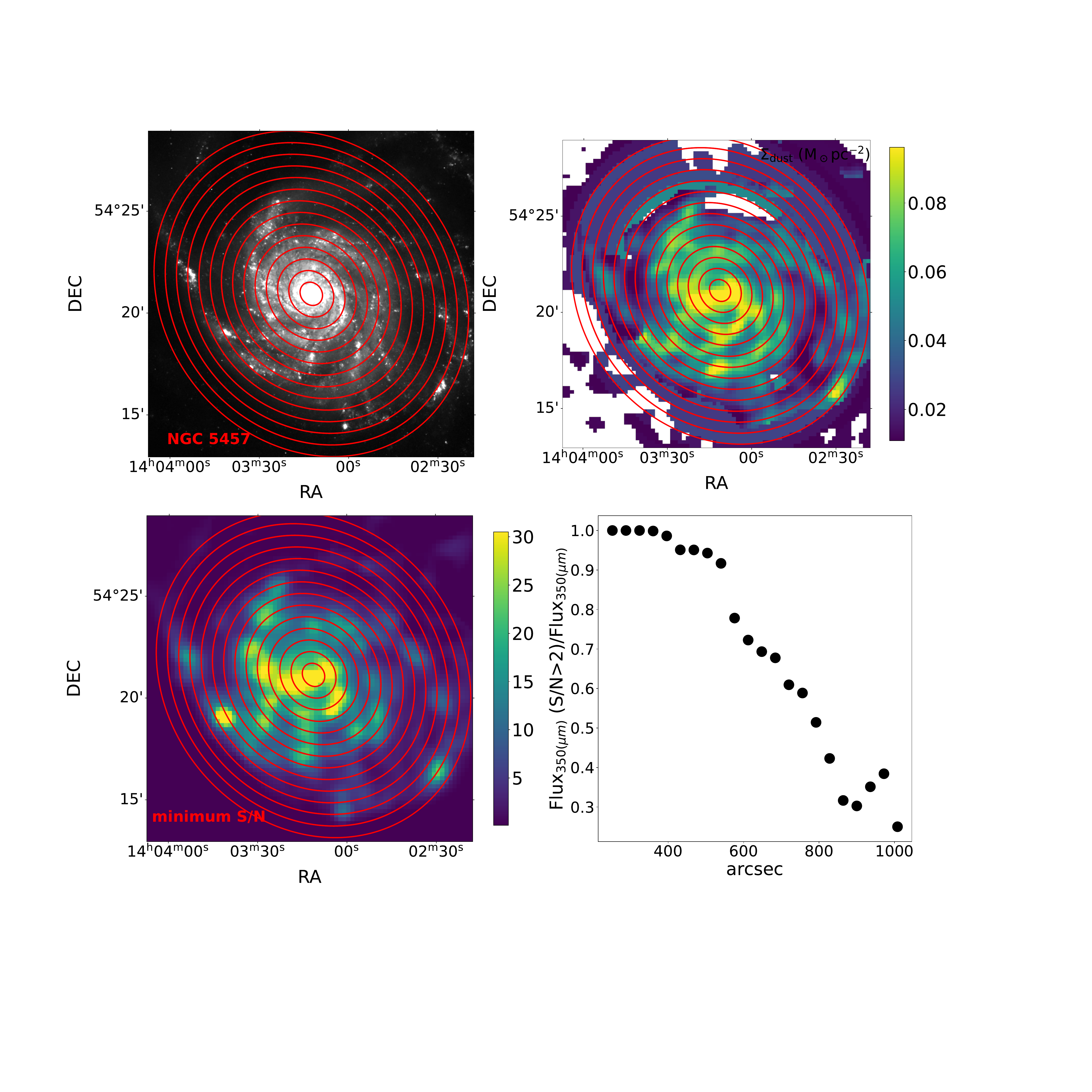}
    \caption{Spatially resolved images of NGC\,5457. Top left: SDSS $g$-band image of NGC\,5457 with concentric annuli of 36\,arcsec width. Top right: Dust mass surface density of the whole disk. Bottom left: S/N map corresponding to the lowest S/N of all the observed fluxes in each pixel. Bottom right: Total 350$\mathrm{\mu m}$ flux of the pixels having a S/N above 2 in the 350\,$\mathrm{\mu m}$ band divided by the total 350$\,\mathrm{\mu m}$ flux of the pixels with low S/N in all bands for each ring.}
    \label{fig:dustmethod_5457}
\end{figure}

We illustrate in Figs.\,\ref{fig:dustmethod_300}, \,\ref{fig:dustmethod_598}, and \,\ref{fig:dustmethod_5457} the methodology used to derive the dust surface density maps for NGC\,300, NGC\,598, and NGC\,5457. 

\subsection{Comparison with the Themis dust model}\label{App:CompThemis}
As we have explained in the main text, the dust model we have used here \citep{1990A&A...237..215D} is a relatively simple one that separates the total population of dust grains into PAHs, VSG and BG. A more sophisticated dust model that tries to incorporate how the individual dust grains respond to the physical properties of the local environment and has a somehow more holistic evolutionary approach is The Heterogenous dust Evolution Model for Interstellar Solids (Themis) \citep{Jones17}. This model separates amorphous carbonaceous dust grains (named as a-C(:H)) into small grains that follows a power-law grain size distribution and large grains with a log-normal distribution. The other main dust grain type is large amorphous silicates with Fe and FeS nano-inclusions. In this model there are no PAHs per se, both aromatic and aliphatic hydrocarbons are part of the same amorphous carbonaceous dust grains. However, the small (a < 1.5\,nm) a-C(:H) grains would correspond to those carrying the mid-IR features in the observed spectrum \citep[see fig.\,1 in][]{Gall22}. We can parameterized the dust grain size distribution of the amorphous carbon grains in such as way that we can retrieve the mass fraction of those grains producing the mid-IR features in Themis dust model, which would correspond to the PAHs in the \citet{1990A&A...237..215D} dust model. In this adaptation of the Themis dust model, a-C(:H) grains smaller than 1.5\,nm are considered as PAHs, larger a-C(:H) grains with a power-law grain size distribution would be considered as the very small grains in  \citet{1990A&A...237..215D} dust model, and large a-C(:H) grains with a log-normal distribution together with the large amorphous silicates would form the large dust grains. With this separation we can then proceed to apply the same fitting methodology as described in Sect. \ref{Sec:dustmaps} and derive the corresponding Themis dust mass map and small to large grain ratio map. The same adapted dust grain size distribution of Themis dust model was used to fit successfully the SED of a large sample of nearby galaxies in \citet{Gall22}. 

\begin{figure}
    \centering
    \includegraphics[width=0.8\linewidth, clip=true, trim=0mm 0mm 3mm 3mm]{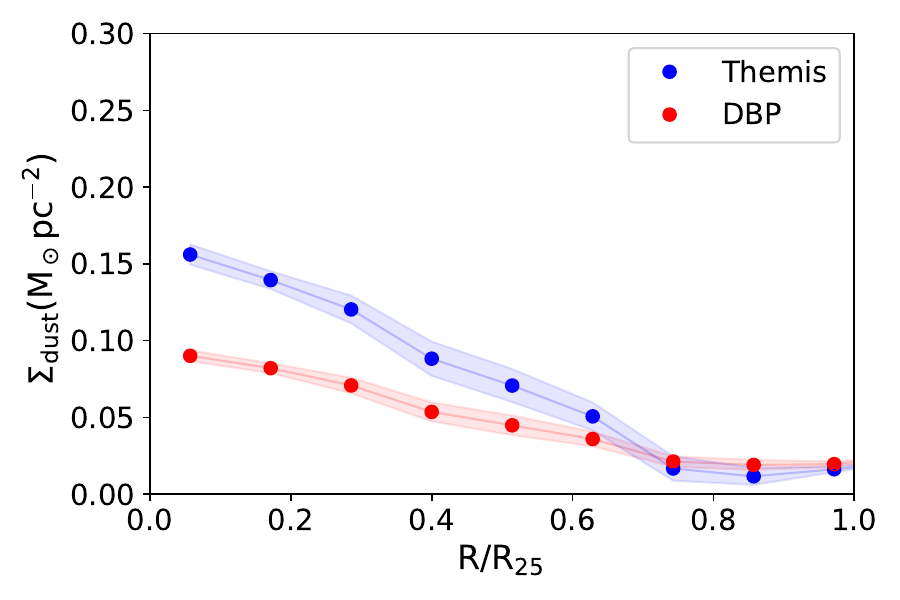}
     \includegraphics[width=0.8\linewidth, clip=true, trim=0mm 0mm 3mm 3mm]{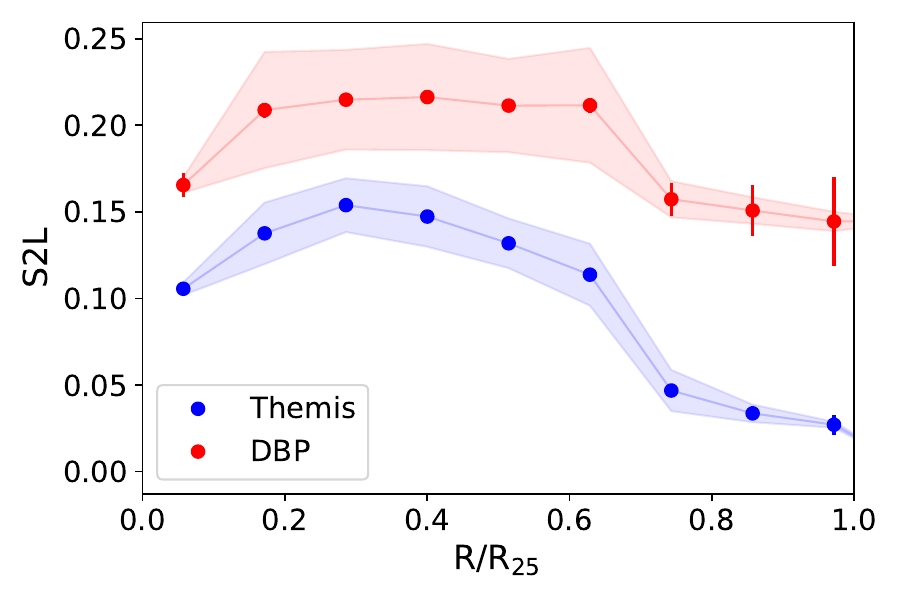}  
    \caption{Comparison between the dust mass surface density (left) and SLR (right) derived in this work with those derived using Themis dust model \citep{Jones17}.}
    \label{fig:themis-comp}
\end{figure}

In Fig.\,\ref{fig:themis-comp} we show the comparison of the radial profiles of the dust mass surface density (left) and the SLR (right) between the dust model used in this work and the adaptation of the more sophisticated Themis dust model we have explained above. The radial trends for the dust mass surface density and SLR are the same. Differences in the dust mass surface densities of the order of 0.05 M$_{\odot}\rm pc^{-2}$ are mainly in the inner part of the disk, with Themis dust model retrieving larger dust mass SD than the model used here. This agrees with the results from \citet{Gall21} who also found that Themis dust model gives larger dust masses for their galaxy sample. In our case the higher dust masses retrieved by Themis dust model can be explained by the different normalizations of the dust extinction coefficient in both models. Themis dust model has a FIR opacity described as $\rm \kappa(\lambda)=6.4\,cm^{-2}g^{-1}\times(250\,\mu m/\lambda)^{1.79}$, while the opacity in \citet{1990A&A...237..215D} dust model ($\rm \kappa(\lambda)=4.8\,cm^{-2}g^{-1}\times(350\,\mu m/\lambda)^{1.79}$), would give a $\rm \kappa_{250\,\mu}=9.4\,cm^{-2}g^{-1}$, a factor 1.5 lower than the one in Themis dust model at 250$\,\mu m$. This factor is similar to ratio of the dust mass SD retrieved by both dust models in the inner part of galaxy.  

\section{Time evolution of the matching}\label{App:Time}
\begin{figure}[!ht]
    \centering
    \includegraphics[width=\linewidth
    , clip=true, trim=5mm 0mm 0mm 0mm
    ]{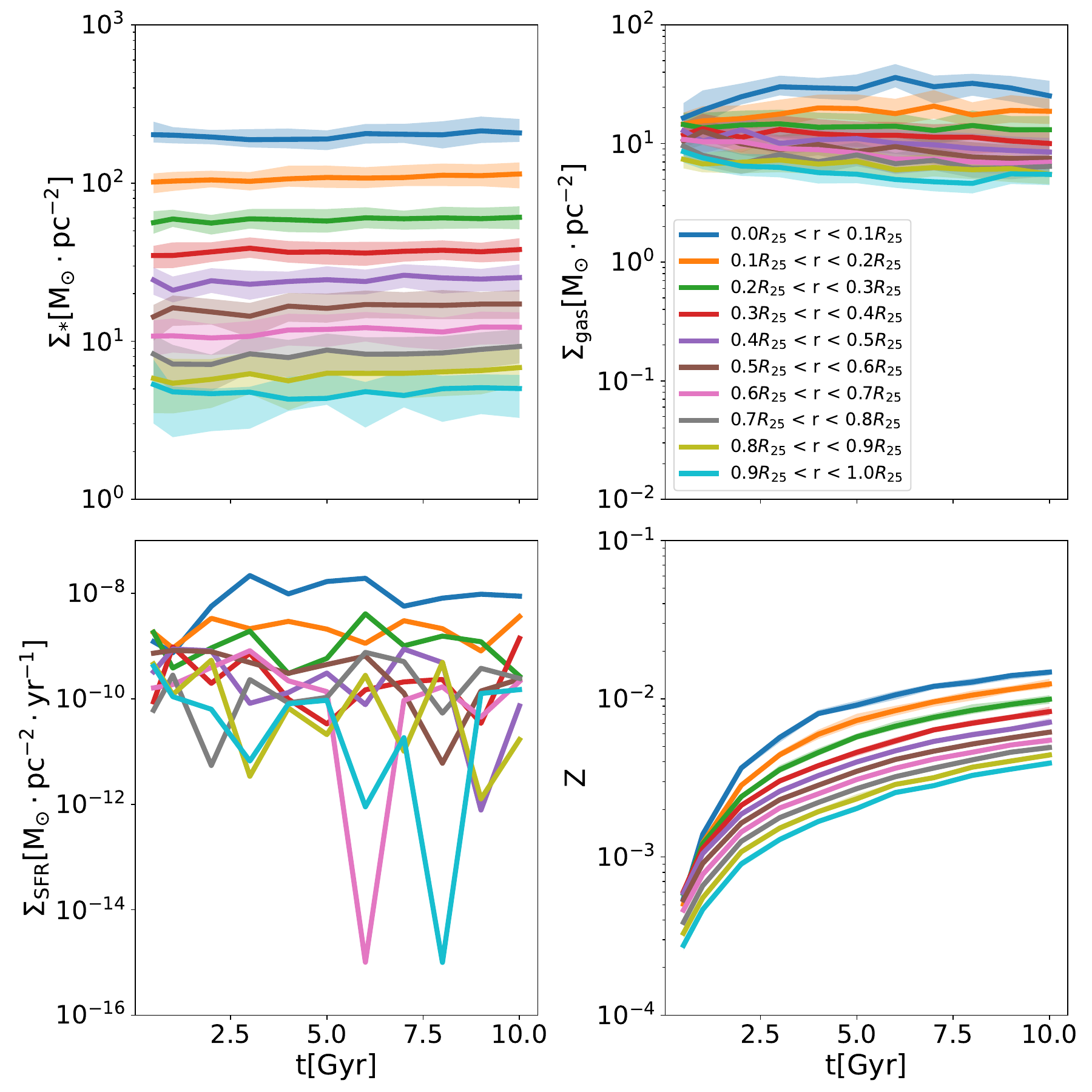}
    \caption{$\mathrm{\Sigma_{*}}$, $\mathrm{\Sigma_{gas}}$, $\mathrm{\Sigma_{SFR}}$, and Z evolving over time in the different rings used for the radial comparison. Only the metallicity, $Z,$ has a strong evolution over time.}
    \label{fig:SGFRZ_1SD}
\end{figure}

In Fig.~\ref{fig:SGFRZ_1SD} we show how the radial gradient of the simulated physical properties used to determine at which time the simulated galaxy agrees with NGC\,628. We notice here that the stellar mass and gas mass remain roughly constant across the entire evolution of the simulation, whereas the SFR shows some scatter due to the stochastic nature of the implementation. The only property that significantly evolves is the metallicity. Therefore, we only look at the Z for the matching. It is worth noting that NGC628 might not have had a constant stellar mass and gas mass throughout its evolution. That might explain why our molecular gas mass fraction and/or SLR do not match between observations and simulations.
\end{document}